\documentclass[twocolumn]{aastex61}

\usepackage{subfigure}
\usepackage{url}
\usepackage{hyperref}
\usepackage{longtable}
\usepackage{natbib}
\usepackage{amsmath}
\usepackage[normalem]{ulem}
\usepackage{bm}
\usepackage{comment}
\usepackage{array}
\newcolumntype{P}[1]{>{\centering\arraybackslash}p{#1}}
\newcolumntype{M}[1]{>{\centering\arraybackslash}m{#1}}
\usepackage{color}

\newcommand{\kep}{{\it Kepler}}
\newcommand{\kt}{{\it K2}}
\newcommand{\TESS}{{\it TESS}}
\newcommand{\Gaia}{{\it Gaia}}

\newcommand{\teff}{$T_{ eff}$}

\newcommand{\logg}{{log(g)}}

\newcommand{\Sph}{{$S_{\textrm{ph}}$}}
\newcommand{\RHK}{{$R'_{\textrm{HK}}$}}

\newcommand{\uw}{Astronomy Department, University of Washington, Seattle, WA, 98195, USA}
\newcommand{\chicago}{Department of Astronomy and Astrophysics, University of
Chicago, 5640 S. Ellis Ave, Chicago, IL 60637, USA}
\newcommand{\sagan}{Sagan Fellow}

\begin{document}
\title{Long Term Photometric Variability in \textit{Kepler} Full Frame
Images: Magnetic Cycles of Sun-Like Stars}

\shorttitle{Magnetic Cycles of Sun-Like Stars} 
\shortauthors{Montet, Tovar, and Foreman-Mackey}

\author{Benjamin~T.~Montet}
\altaffiliation{\sagan}
\affiliation{\chicago}

\author{Guadalupe~Tovar}
\affiliation{\uw}

\author{Daniel~Foreman-Mackey}
\altaffiliation{\sagan}
\affiliation{\uw}

\correspondingauthor{Benjamin~T.~Montet}
\email{bmontet@uchicago.edu}

%@arxiver{f8a.pdf,f5.pdf,f2b.pdf}
%\date{\today, \currenttime}

\begin{abstract}
Photometry from the \kep\ mission is optimized to detect small, short duration
signals like planet 
transits at the expense of long-term trends.
This long-term variability can be recovered in photometry from the Full Frame Images (FFIs), a set of
calibration data collected approximately monthly during the \kep\ mission. 
Here, we present \textit{f3}, an open-source package to perform
photometry on the \kep\ FFIs in order to detect changes in the brightness of stars in the \kep\ field
of view over long time baselines.
We apply this package to a sample of 4,000 Sun-like stars with measured rotation periods.
We find
$\approx 10\%$ of these targets have long-term variability in their observed flux. For the majority of
targets we identify the luminosity variations are either correlated
or anticorrelated with the short-term variability due to starspots on the stellar surface.
We find a transition between anticorrelated (starspot-dominated) variability and correlated
(facula-dominated) variability between rotation periods of 15 and 25 days, suggesting the transition 
between the two modes
is complete for stars at the age of the Sun.
We also identify a sample of stars with apparently complete cycles, as well as a collection of
short-period binaries with extreme photometric variation over the \kep\ mission.
\end{abstract}


\section{Introduction}
\label{sec:intro}

For more than two millenia, observations of the Sun have shown the presence of spots 
on its surface. Regular telescopic observations of these spots date back more than four centuries 
\citep{Galilei13}.
These spots have been known to vary on an 11-year timescale \citep{Schwabe44}, with more than two dozen 
of these cycles now observed \citep[][and references therein]{Hathaway15}.
This starspot cycle is likely produced by the solar dynamo \citep{Charbonneau10}.
While the most obvious effect of the solar cycle is variation in the location and number of sunspots,
it also correlates with the occurrence of coronal mass ejections \citep{Gosling93} and a
change in the total solar irradiance \citep{Frohlich98}.
Over the solar cycle, the Sun changes in brightness by approximately 0.1\% in the optical,
with times of increased starspot activity corresponding to an increase in luminosity
due to the bright faculae that typically surround darker sunspots \citep{Frohlich04}.

More recently, similar behavior has been seen on other Sun-like stars.
Spectroscopic observations at Mt. Wilson showed main sequence stars from spectral types F5 to M2 
have chromospheric variations in their atmospheres, providing evidence for analogs of the
solar cycle \citep{Wilson78, Baliunas95}.
The level of chromospheric variation has been shown to correlate with stellar rotation 
\citep{Noyes84, Saar99}.
Additional observations have shown that Sun-like stars also have photometric
variations, with more rapidly rotating stars having larger variations in flux
\citep{Lockwood97, Radick98}.
Magnetic cycles have also been observed in time-series radial velocity (RV) planet searches,
as magnetic activity affects the convective blueshift, leading to an apparent RV shift with the period
of the magnetic cycle \citep{Endl16}.

Observations of different stars at different ages have provided opportunities to detect evolution of the
stellar dynamo. 
Presumably older stars with slower rotation rates have longer magnetic cycles with lower amplitudes
\citep{Noyes84, Baliunas95}. 
Observations of rotation periods and activity cycles have suggested there are two branches of Sun-like
stars, an ``active" branch with longer magnetic cycles \deleted{and higher chromospheric activity} and an ``inactive'' branch with shorter cycles \added{for
a given rotation period} \deleted{and lower 
chromospheric activity}
\citep{Bohm-Vitense07}. Interestingly, some stars have shown behavior consistent with both branches.
The Sun appears to fall between these two branches, both in the length of its magnetic cycle
and its observed chromospheric activity, suggesting the solar dynamo may be in transition
\citep{Metcalfe16}. An additional transition has been proposed, with a transition from complex to smooth
magnetic cycles suggested at an age between 2 and 3
Gyr \citep{Olah16}.
This transition has also been suggested to lead to a change from starspot-dominated photometric
variations to facula-dominated variations, although the details of any particular star depend on 
the star's inclination and latitudinal distribution of spots \citep{Shapiro14}.

Our understanding of the solar dynamo and its potential transition is limited by the small number of
stars with detailed observations of their magnetic activity.
Spectroscopic observations are expensive, photometric variations at the sub-mag level can be small and
hard to detect from the
ground for all but the brightest, nearest stars \citep{TregloanReed13}. 
\added{Asteroseismic observations can be used to measure global changes in stellar oscillation modes
due to changes in the overall stellar activity level, which have been used to detect stellar activity cycles
with both CoRoT and \kep\ \citep{Garcia10, Salabert16a}.}
However, asteroseismic observations require high-precision
photometry at a \added{fast enough cadence to resolve the modes of pulsation. For Sun-like stars, long cadence 
observations are insufficient; short cadence is required but only present for a small fraction of these targets.}  \citep{Gilliland10}. 
Even over small spatial scales, ISM absorption can bias spectroscopic stellar activity observations: 
a star 100 pc away can have a bias in its \RHK\ value by as much as 0.1 dex due to interstellar Ca$_{\textrm{II}}$ \citep{Fossati17}.
In total, there are only approximately 100 Sun-like stars with measured activity cycles
\citep[e.g.][]{Hall09}.

Photometric tracers of magnetic activity can be obtained from space-based transit surveys, like
\kep. Launched in 2009, \kep\ targeted nearly 200,000 stars across 115 square degrees of the sky,
and was optimized to detect small changes in brightness due to the presence of planets transiting the disk
of each star \citep{Borucki10}. 
The primary data product, ``long cadence photometry'' includes observations of each star at approximately
30-minute intervals over a span of four years \citep{Jenkins10}.
These data are ideal for detecting short-term changes in the brightnesses of these stars, such as 
starspots rotating into and out of view \citep{Nielsen13, McQuillan14}, stellar flares 
\citep{Hawley14, Davenport16}, and asteroseismic
pulsations \citep{Huber13, Aguirre15}.
\citet{Mathur14} have observed changes in the starspot variability of F stars in \kep\ data as a proxy
for magnetic activity, an approach that \citet{Reinhold17} have extended across the main sequence.  
Changes in the \deleted{phase} \added{frequency} shift of asteroseismic modes have also provided an 
opportunity to probe magnetic activity \citep{Salabert16a}.
However, signals on longer timescales ($\sim 50$ days or longer) such as brightness variations on the 
timescale of magnetic cycles,
are both intentionally removed by the data processing pipeline and overwhelmed by instrumental systematics
due to the small aperture sizes relative to the size of the telescope PSF \citep{Gilliland11}.

Long-term brightness variations can be recovered through the Full Frame Images (FFIs), a set of
calibration data obtained approximately monthly during the \kep\ mission. 
These observations are the only times the entire detector, encompassing all 4.5 million stars in the
field of view, is downloaded.
FFI data have been used to help confirm signals in long cadence data and for visualization purposes
\citep{Lehmann12, Jenkins15, Gaidos16}. 
On their own, they contain the entire point spread function (PSF) for each star and a large number of reference stars nearby
on the detector, enabling long-term brightness variations removed from the long cadence data to be
recovered \citep{Montet16b}.

In this paper, we develop a method to recover long-term brightness variations for stars in the \kep\
field from FFI data, building on previous work \citep{Montet16b}. 
In Section 2, we explain in detail how to measure photometry from
the FFIs using our publicly available code, then use it to explore long-term flux variations in a sample
of more than 3,000 Sun-like stars with measured rotation periods. 
In Section 3, we present the results of our search and verify we are observing astrophysical variability
from the target stars themselves. In section 4, we discuss particularly interesting systems, 
and present evidence for a \deleted{transition in the stellar dynamo at an age of 2-3 Gyr} \added{shift in the relation between
short- and long-term photometric variability at a rotation period of $\approx 25$ days},
rather than at Solar ages. In section 5, we conclude with a discussion of future prospects of extending
our sample with data from other space missions.

\section{Data Analysis}
\subsection{Sample of Sun-like Stars}
\label{sec:targets}

In this work, our goal is to consider Sun-like stars.
While spectroscopic surveys of the \kep\ field are underway \citep{DeCat15, Guo17}, these surveys generally
target the bright stars in the field. 
Photometric surveys are more complete \citep{Brown11, Huber14}, but provide larger uncertainties on 
the physical parameters of each individual star.
As our sample is large and dominated by faint stars, we build a sample of stars using established
photometric stellar parameters.

\citet{McQuillan14} detect rotation periods in 34,000 main sequence stars in the \kep\ field and
report non-detections on nearly 100,000 additional stars.
Since stellar rotation periods are correlated with the age of the star 
\citep{Barnes07, Mamajek08, Angus15}, these stars provide us the opportunity to not only search for
activity signatures, but to understand their evolution with stellar age.

We select all stars in the \citet{McQuillan14} catalog with measured rotation periods, estimated \teff\
within 150 K of the Sun, and \logg $> 4.2$. This particular catalog uses the stellar parameters of \citet{Brown11}. While more recent publications have updated stellar parameters \citep{Huber14},
the KIC is more homogeneous.
\citet{Pinsonneault12} have shown that this catalog systematically underestimates effective temperatures
by approximately 200 K, which, combined with random uncertainties of approximately 100 K, suggests our sample
should largely be stars with spectral types F7 to G4.

\begin{figure*}[htbp!]
\centerline{\includegraphics[width=0.95\textwidth, clip=True]{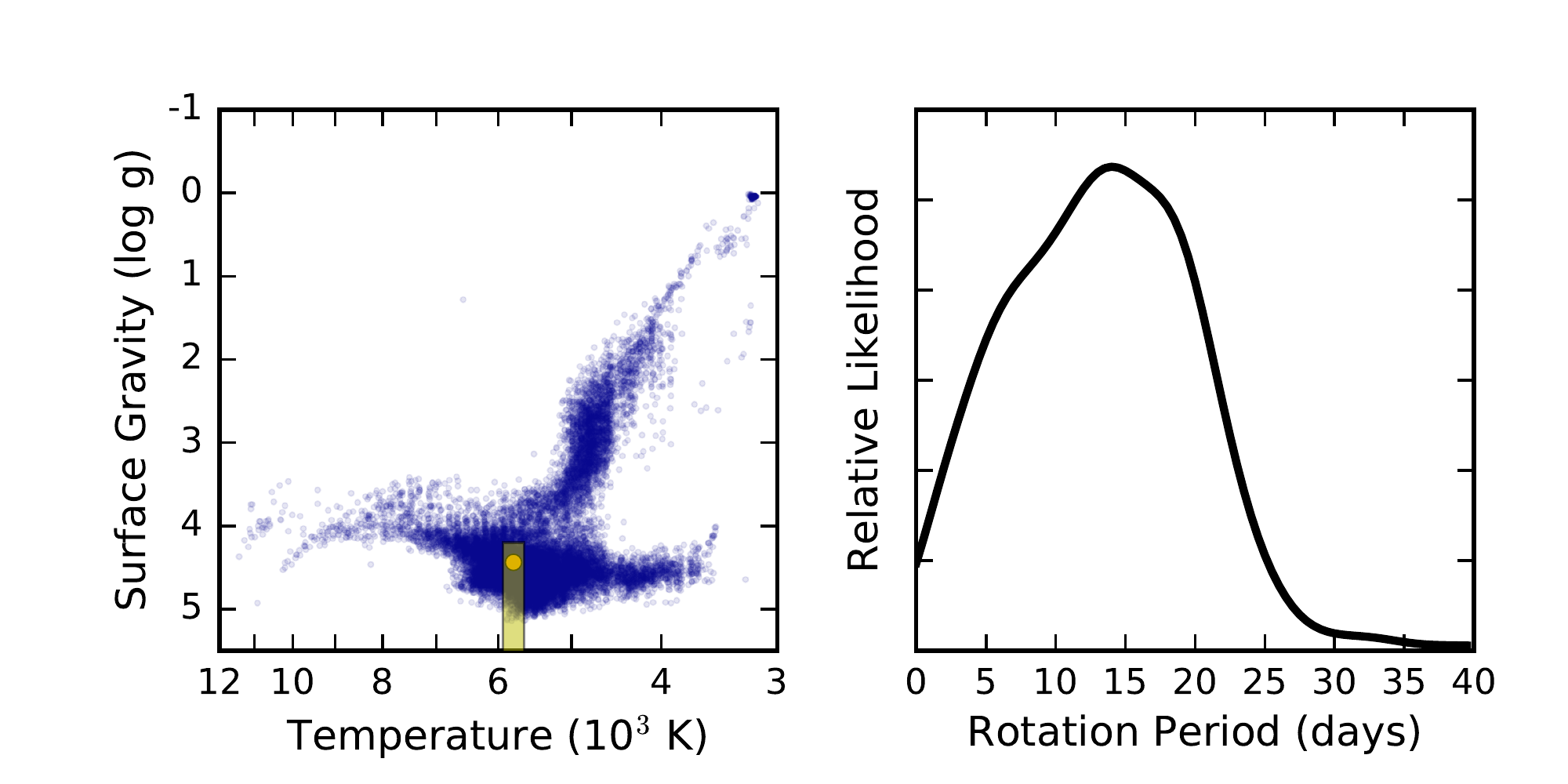}}
\caption{(Left) KIC-reported effective temperatures and densities for stars observed during the primary
\kep\ mission.
The region we draw our target sample from is shaded in yellow. For illustration, the Sun is shown as a 
yellow dot. (Right) Kernel density estimation of the distribution of rotation periods in our sample of 
stars. The typical star is younger than the Sun: the distribution peaks at 14 days, while the Sun's 
rotation period is 25.38 days \citep{Soderblom10}.
}
\label{fig:params}
\end{figure*}

The \citet{Bastien16} catalog of stellar 
\logg\ inferred from the \kep\ data is limited to bright \kep\ stars with $K_p < 13.5$, while the majority
of stars in our sample are fainter than that value, limiting our ability to intentionally exclude evolved
stars. 
However,
the requirement that each star has a measured rotation period should also ensure that the number of
evolved stars with very slow rotation periods is minimal. 
The distribution of stellar effective temperatures and
rotation periods is shown in Fig. \ref{fig:params}.

These cuts provide us with a total of 4876 stars. We additionally remove all stars that fall within
10 pixels of the detector edge, which would complicate aperture photometry calculations.
We then select a similar number of stars with the same stellar parameter cuts, but no detected rotation
periods to use as a control sample. These stars have similar colors and brightnesses, but are older,
less active, and possibly evolved, so we should expect to see less variability in these stars. 
Differences between these two samples ensure that our observations are dominated by astrophysical
information rather than underlying, unknown instrumental systematics.
Once we have a list of KIC identifiers associated with the stars in the sample, we can perform photometry on
each of these to search for long-term brightness variations.

\subsection{Full Frame Images}

\subsubsection{Data Collection}

The observations in this paper are taken from the 53 FFIs collected
by \kep\ during its primary mission. 
These data represent the only publicly available data of simultaneous 
observations of the entire \kep\ field of view.
Eight FFIs, called the ``Golden FFIs,'' were obtained over 34 hours during commissioning of the 
spacecraft before
the start of the primary mission. 
Additional FFIs were obtained approximately once per month during the primary mission,
immediately before the spacecraft turned towards Earth to transmit data.
Two of these observations were missed because the spacecraft entered a safe mode state
before data downlink.
Additionally, on two occasions two FFIs were collected in succession. 
In one case the telescope was mispointed by 15 arcseconds, or approximately four pixels. 
We do not include this image in our analysis. On the second occasion, two acceptable images were obtained.

\added{All FFIs have integration times of 29.4 minutes, identical to that of a standard long cadence frame during
the \kep\ and \kt\ missions. The observations are obtained and calibrated following the same procedure as that of
long cadence photometry, but applied to every pixel rather than a small number of postage stamps. Therefore, we
expect that the photometry from all FFIs to be uniform, and noise properties for individual pixels should be similar
to those observed in standard \kep\ photometry.} 

Each year during the primary mission, \kep\ rolled by 90 degrees four times in order to keep its
solar panels pointed at the Sun. 
The pointing is stable so that, after four rolls, each star falls back on the original
pixel.
As a result, the centroid of each star falls on four different pixels over the mission, each with
its own unknown response function. 
The data from these four orientations can and should be treated as four different sets of
observations, each with its own noise properties and potential systematics.

All FFIs are publicly available and contain observations of approximately 4.5 million objects
in the telescope's field of view. 
We downloaded all FFIs from the Mikulski Archive for Space Telescopes (MAST).
Each image is already calibrated through a process similar to the \kep\ pixel calibration
pipeline (``CAL''), with the \added{bias level}, dark \added{current}, and smear
\added{level, the result of stars illuminating the CCD detector during readout
causing apparent streaks on the detector,} estimated and corrected \citep{Jenkins10, Quintana10}.

\subsubsection{Data Reduction}
\label{sec:pipeline}

In this work, we develop time-series photometry for each of the target stars as described in
Section \ref{sec:targets}. The method used is broadly similar to that of \citet{Montet16b},
but is more completely automated and achieves a typical precision higher than achieved by
those authors by a factor of 2-3, depending on the target star and field.
For all stars in our sample, we identify the row and column on the detector on which our
target star appears, \added{using data from the Kepler Input Catalog, which contains position
information for each star at arcsecond precision, more than sufficient for our purposes}.
We then select 300$\times$300 pixel (20 arcmin) ``postcard'' regions around each star (Fig. 2).
In most cases, these regions are centered on the target star; when the target star falls near
the edge of a detector they are shifted to ensure we always consider an equal-sized region.
We build a master postcard for each field by summing over all 52 usable FFIs, allowing for
shifts of up to one pixel in either direction \added{in each individual frame} to account for apparent pointing variations
induced by differential velocity aberration \citep{Haas10}.
With this image, we identify the 250 brightest sources on the postcard corresponding roughly to stars
brighter than $K_p = 16.5$\footnote{This value depends on the density of stars in the field, which itself
depends on the location of the star on the detector.}.
This selection will include our target star, while we draw our photometric comparison stars
from the collection of other nearby stars that we select, typically around 200 in 
number.

\begin{figure*}[!tbp]
  \centering
  \begin{minipage}[b]{0.48\textwidth}
    \includegraphics[width=\textwidth, trim={1.88in 3.035in 1.5in 1.5in}, clip=true]{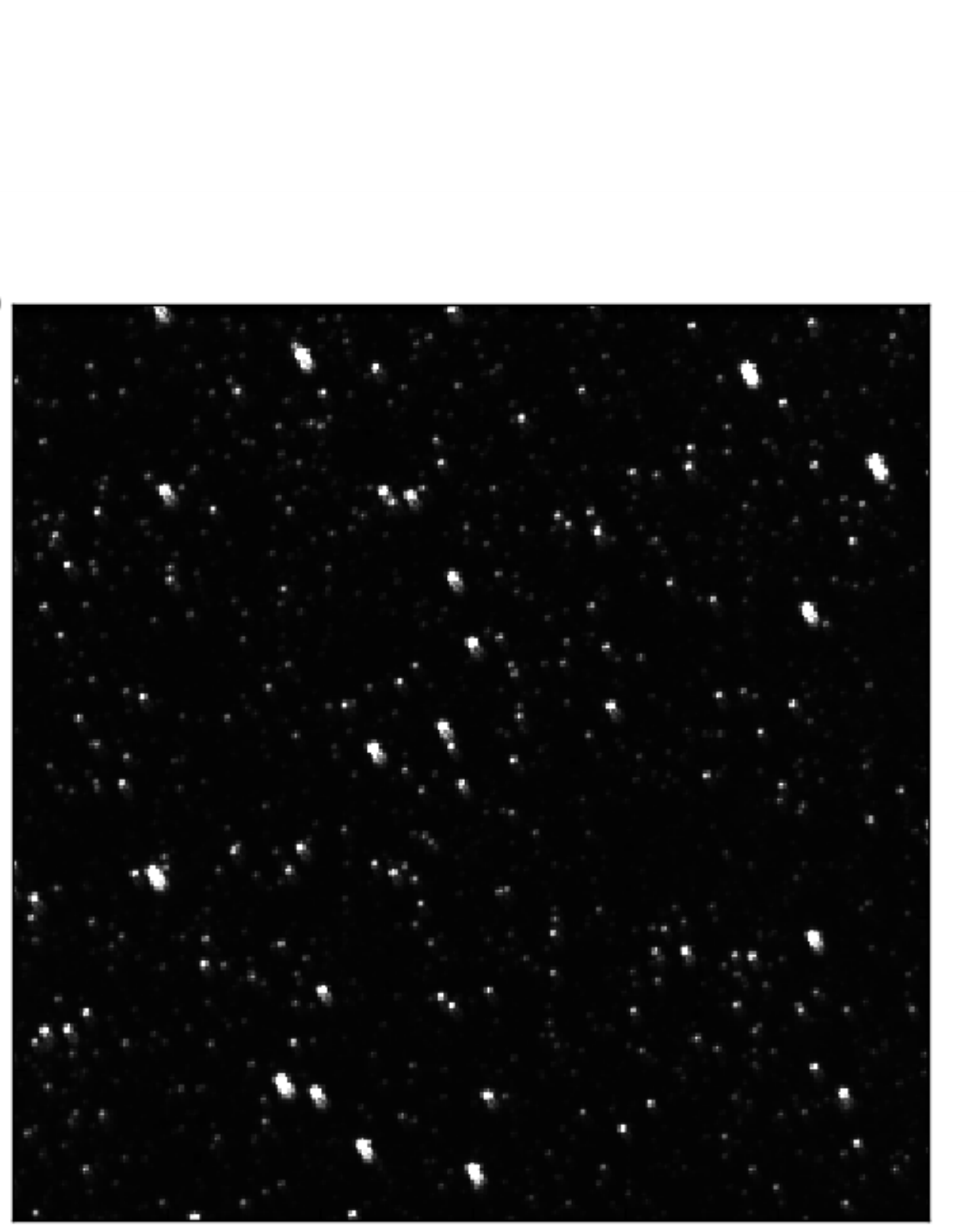}
  \end{minipage}
  \hfill
  \begin{minipage}[b]{0.48\textwidth}
    \includegraphics[width=\textwidth, trim={1.88in 3.035in 1.5in 1.5in}, clip=true]{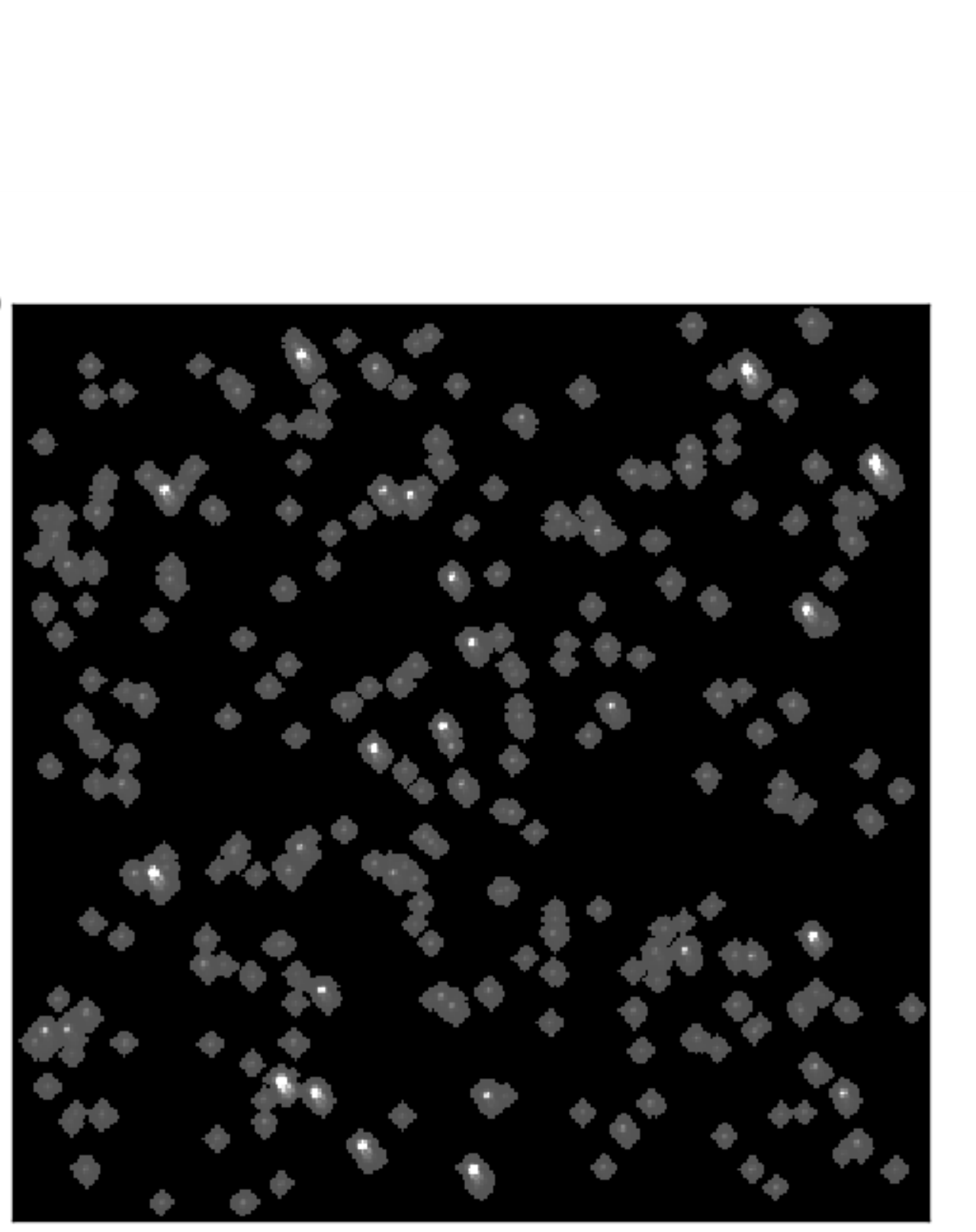}
  \end{minipage}
  \label{fig:postcard}
  \caption{(Left) Typical postcard region of the detector from which reference stars are drawn.
  The target star, in this case KIC 8462852, is directly in the center of the image. (Right) 
  The same postcard, with apertures drawn over reference stars. The 250 brightest targets are
  considered; targets with apertures that touch the edge of the postcard are removed; targets
  which apertures overlap are combined and treated as one reference star.}
\end{figure*}

Once we have identified targets, the next step is to measure flux values for all targeted stars
in each epoch. 
One possibility is to employ PSF modeling.
Detailed PSF modeling is challenging with \kep\ as the underlying flat field is poorly understood.
PSF modeling has been shown to work for faint stars in crowded \textit{K2} fields, but underperforms simple aperture photometry for bright ($K_p \lesssim 15.5$), isolated stars \citep{Libralato16}.
We also explore model PSFs as the sum of three Gaussian functions, conditioned on all target stars
in the postcard, but find that the uncertainties in the model fit are larger than those that we
are able to achieve in simple aperture photometry. Therefore, we rely on aperture photometry
in this work, leaving a detailed exploration of the possibility of PSF modeling for future efforts.

Aperture photometry requires accurate apertures that are large enough to capture the extent of the stellar PSF but small enough to only include the target star.
To capture the \kep\ PSF, the ideal aperture is asymmetric, especially near the edges of the
detector \citep{Bryson10}.
Here, we use the computer vision library \textit{mahotas} to draw appropriate apertures around
our targets and separate them from the background \citep{mahotas}.
We first choose all connected pixels with flux values larger than 1.5\% of the brightest pixel's flux,
then extend the aperture to include a border of three pixels outside this region.
\added{This selection criteria creates a region of the detector large enough to encapsulate the entire
PSF but small enough to isolate only the star in question.
Any aperture where both of these criteria are true enable the light curve to be recovered.}
In cases where multiple comparison stars have apertures that collide with each other, we combine
their apertures into one and treat them as one single target.

We visually inspect each aperture for our science targets to ensure they contain only one star, 
manually removing stars with nearby, bright companions that encroach on the primary star's 
PSF: due to the large PSFs on the detector, this removes approximately 10\% of all possible stars
from our analysis.
Once our apertures are determined, we measure fluxes for all stars by summing over all pixels in
each aperture.

\begin{figure}[htbp!]
\centerline{\includegraphics[width=0.45\textwidth]{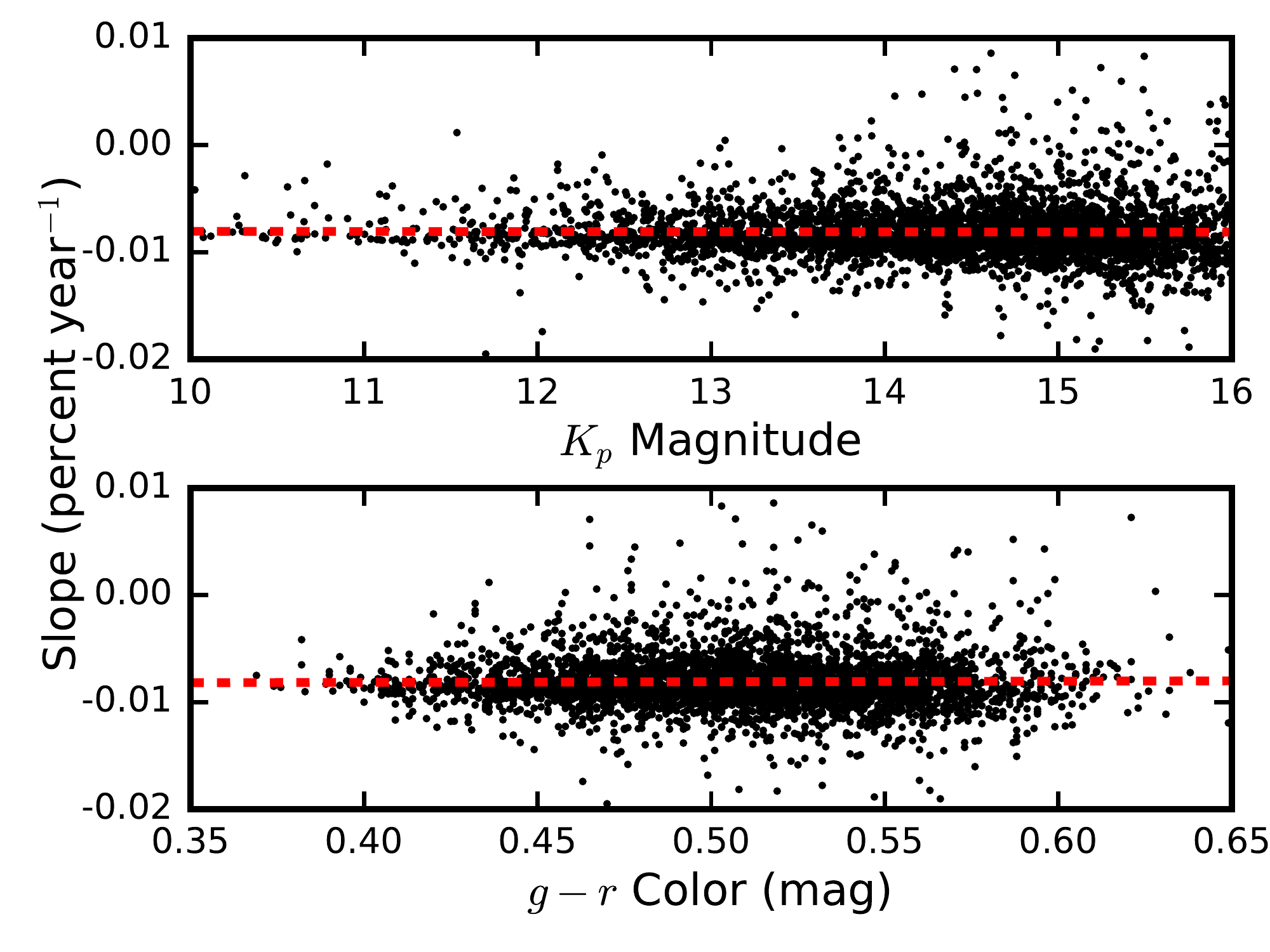}}
\caption{Observed change in brightness in time for all stars in our sample as a function of $K_p$ magnitude
and $g-r$ color, an instrumental effect observed in the FFI photometry and standard \kep\ data. There is a slow decay which
is instrumental and shared by all stars on the same part of the same detector. Red dashed lines represent
a linear fit to the slope as a function of magnitude or color; we find both are consistent with zero,
suggesting there is not an instrumental systematic which specifically affects saturated stars or particularly red or blue wavelengths. We therefore use all nearby stars as potential
reference stars, regardless of their stellar parameters.
}
\label{fig:colormag}
\end{figure}

For all stars, we notice a slow decrease in the observed flux. \added{If this decrease were a function of
color or magnitude, it would limit our ability to use any nearby star as a potential calibrating reference star.
Fortunately, this does not appear to be the case:} the magnitude of this decrease \added{appears to be the
same for all stars on the same detector} \deleted{does not appear to not be a function
of color or apparent magnitude} (Figure \ref{fig:colormag}). 
Therefore, we consider all \added{nearby} stars to \added{share similar photometric systematics with our
target star, making them} \deleted{be} potentially acceptable reference stars\deleted{, regardless of their
spectral type or magnitude}.

We invoke two methods to use the reference stars to measure the true long-term photometric
behavior of each of our target stars. As a first pass, we simply eliminate all possible reference
stars with excess variability ($2.5\sigma$ from the mean) in more than 10\% of frames. This
should ensure intrinsically variable stars and stars with poorly drawn apertures are removed
from the sample of reference stars. For all images in a particular orientation, we then sum
the flux of all reference stars and divide the observed flux from our target star by this value.
We divide each observed flux value by the median flux value for the target star from all epochs
in that orientation, effectively eliminating possible effects caused by variations in the flat
field between detectors used at different orientations. 
This method is fast and enables us to quickly measure the
brightness of each star to look for candidate active stars.

For each of the candidate active stars, we then employ a probabilistic approach---similar in spirit to the ``ubercalibration'' method used in the \emph{Sloan Digital Sky Survey} pipeline \citep{Padmanabhan08}---which enables us to 
measure the brightness of our target star and estimate an uncertainty on each observation.
The full details of this method are given in Appendix~\ref{sect:probcal}, but we summarize the
important points here.
We model the observed light curves of an ensemble of stars as noisy
measurements of intrinsically variable time series observed with varying pixel
responses and photometric zero points.
We make this model tractable by assuming that nearby stars with similar
properties sample a constant photometric zero point and
that the pixel response for a particular source is constant in each season of
observation.
We fit for a distribution of zero points for each exposure, and the pixel
responses and amplitude of intrinsic variability for each target by maximizing
the likelihood function defined by this model (Equation~\ref{eq:likelihood}).
We use this maximum likelihood model to compute the de-trended light
curve and its uncertainties.

The end result is time series photometry for all of our target stars, conditioned on the 
observations of nearby reference stars on the detector, with underlying systematics between
detectors accounted for. These light curves are considerably more sparsely sampled than \kep's
long cadence data, but allow for long-term trends which are overwhelmed by instrumental systematics
in the long cadence data to be recovered. 
We make the code underlying FFI photometry code publicly available for community use and additional
development as the \textit{f3} (\textit{Full Frame Fotometry}) package\footnote{https://github.com/benmontet/f3}.

\added{The uncertainties on each data point are heteroskedastic. We describe the method through which 
uncertainties are estimated in Appendix~\ref{sect:probcal}. For each star, the photon noise observed in
the primary \kep\ mission is not the dominant source of uncertainty. Over short timescales, similar to a
typical transit duration, this is true, but on longer timescales in the \kep\ data changes in the detector
 properties, such as thermal variations and detector degradation,
overwhelm the photon noise: this is evident from the long-term trends seen in standard \kep\ data
over a quarter. 
Indeed, inspecting our photometry over the ``Golden FFIs'' collected across two days, over which the pointing
is stable, shows lower variability than the point-to-point scatter across monthly observations, similar to what
would be observed if, for example, one selected monthly points from a long cadence light curve.
Due to uncertainties in the flat field and detector properties, combined with the sparse sampling of the FFIs, 
our photometric uncertainties on any 
individual point are considerably
larger than those from relative, detrended photometry from the \kep\ mission. For an upcoming mission like
\textit{TESS} \citep{Ricker14}, with an FFI collected every 30 minutes, it will be possible to simultaneously 
model the detector properties and the photometry, similar to what is done in \textit{K2} fields \citep[e.g.][]{Luger16},
enabling a similar study to be accomplished but
recovering a similar photometric precision to that achievable on the primary target stars.}

\subsubsection{Potential systematics}

\begin{figure}[htbp!]
\centerline{\includegraphics[width=0.45\textwidth]{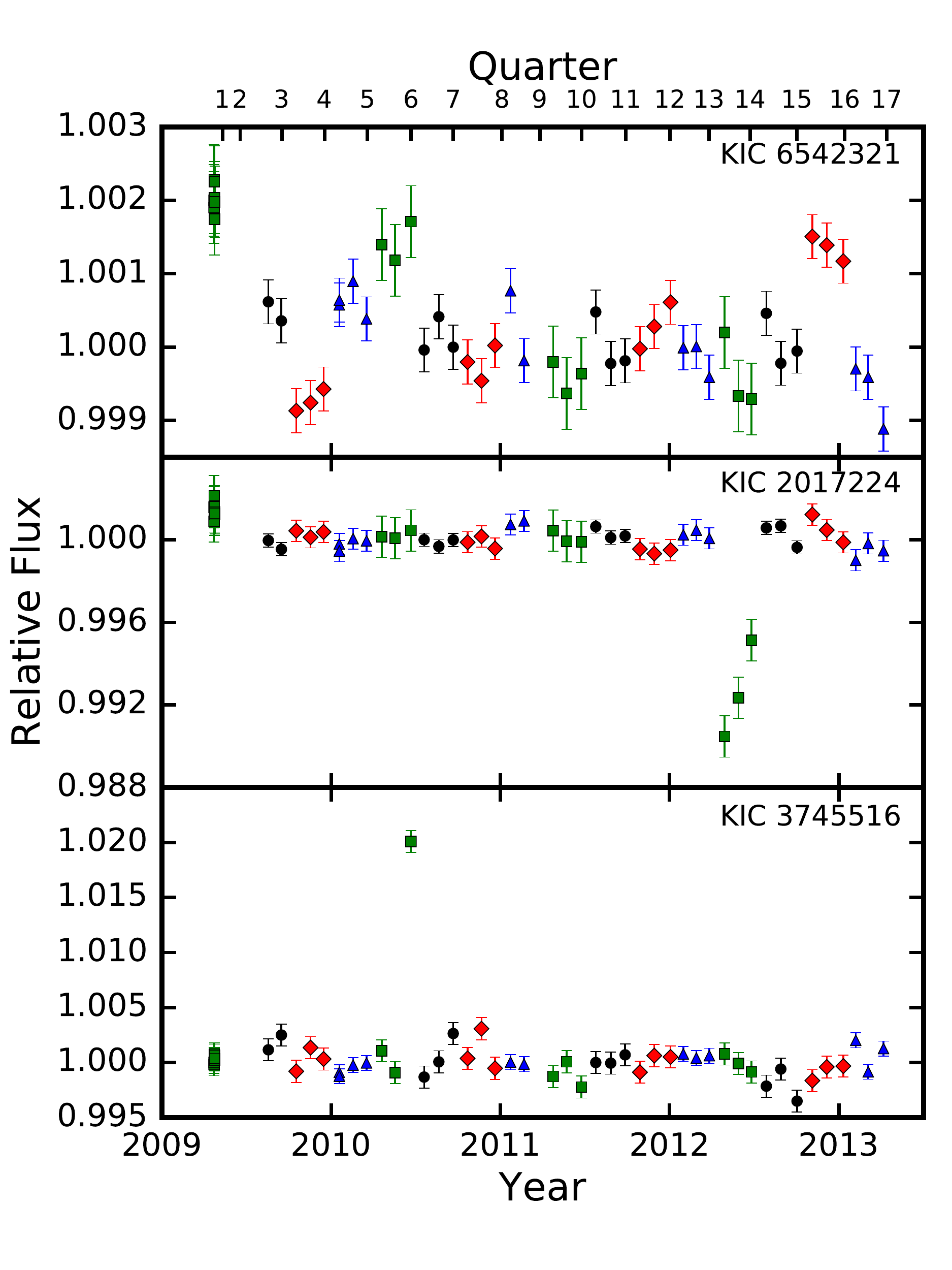}}
\caption{FFI time series photometry for three stars displaying the most common systematics observed in the FFI data. The green squares, blue triangles, black circles, and red diamonds correspond to the four 
orientations of the telescope during one year of observations. (Top) KIC 6542321, a star with
a proper motion of 26 mas yr$^{-1}$. During the mission the star samples different portions of the intrapixel flat field, leading to 
a drift in the flux which is different in the different telescope orientations, and therefore clearly
not astrophysical.
(Middle) KIC 2017224, which has one of its four sets of pixels undergo a sudden pixel sensitivity dropout.
The pixel recovers during Quarter 13, and no other orientations are affected.
(Bottom) KIC 3745516, which is affected by a cosmic ray during Quarter 5. Because our apertures are
significantly larger than those used in long cadence photometry, our data are affected by cosmic rays at a
higher frequency.
}
\label{fig:systematics}
\end{figure}

While they do enable accurate measurements of long-term flux variations on target stars in the field
of view, FFI photmetry also comes with its own set of systematics. 
Many of these are familiar to users of \kep\ and \kt\ data products, although they can manifest 
themselves in ways that may not be familiar.
We show examples of each of these systematics in Fig. \ref{fig:systematics}.

The most common systematic is a Sudden Pixel Sensitivity Dropout (SPSD). In these events, a cosmic
ray hit causes a particular pixel to immediately decrease in sensitivity, leading to a decrease in the
observed flux on that pixel without changing its neighbors \citep{Smith12}. 
Pixels inflicted with SPSDs may recover after hours or days, while others remain degraded through the
remainder of the mission.

SPSDs are visible in \kep\ long cadence light curves of many stars. Since larger apertures are
required to ensure accurate FFI photometry, there are more possibilities to have
pixels affected by an SPSD than in typical \kep\ data. SPSDs can be separated from true astrophysical
variability in the FFIs because changes will manifest themselves in only one of the four telescope
orientations, while the others will appear unchanged.
Difficulties remain in separating astrophysical variability in the first or last quarter of observations from instrumental effects: this is especially true for the start of the mission, as the 
``Golden FFIs'' are separated from the first FFI in the primary mission by 115 days.

The second most common systematic is apparent brightness variations caused by changes in the position
of the star over the underlying unknown flat field. 
\kt\ observations have shown that there are significant interpixel and intrapixel variations in the 
flat field across the detector \citep{VanCleve16}. 
Some stars in the \kep\ field of view can have proper motions as large as 0.1 pixels year$^{-1}$,
and differential velocity aberrations can cause periodic motions at the pixel level.
Flat field variations can then manifest themselves as a long-term trend or periodic behavior with a
period equal to that of the \kep\ telescope, corresponding to each of those causes.
The position of the star on each pixel at the start of the mission is effectively random, and the
flat field between each orientation is different, so these can be separated from astrophysical
variability: if the effect is instrumental,  the variations look different in each of the four orientations and are correlated with the position of the star on the detector.

Additionally, cosmic rays can occasionally cause short brightness increases on a single pixel in a
single FFI. Like SPSDs, these are somewhat more common than are seen in typical long cadence light curves
since we require more pixels in each aperture. However, as these typically only affect a single FFI, while
brightness variations from stellar activity manifest themselves on longer timescales,
they generally do not affect our results significantly.

We visually inspect each of our light curves for each of these effects. 
We calculate the centroids of light in each of our apertures manually, summing the observed flux in
each pixel. As we also verify there is not significant background contamination in our apertures, we
do not expect any significant dilution due to a large number of high proper motion background or
foreground stars
significantly biasing our results.

\subsection{Information from long cadence light curves}

All of the stars in our target sample have measured rotation periods, so by definition they are also
observed at long cadence in the \kep\ mission. 
We use this information both to improve our photometry and to gain extra information in our analysis
beyond a measurement of stellar rotation periods.

First, we use \replaced{them}{the Pre-search Data Conditioning (PDC) versions of the \kep\ light curves 
\citep{Stumpe12} }
to apply a correction to the observed FFI photometry for the distribution of
starspots at the time of the observation. Each FFI gives us an instantaneous observation of the brightness
of the star, but as we know the time of the observation we can infer from the long cadence light curve
whether that star is in a particularly bright or dim orientation due to the instantaneous distribution 
of starspots. 
By comparing the PDC light curve at the last long cadence observation of each month, one hour before the
FFI observation, compared to the average flux value for that quarter, we can correct effects caused by 
one or a small number of starspots that may cause an aliasing effect with the monthly FFI cadence.
We produce light curves with and without this correction to ensure that we do not induce any false 
signals of our own with this correction.
Unfortunately, the vast majority of our reference stars do not have long cadence photometry, so we do not
have the information to build a global solution that accounts for starspots on all reference stars.
Instead, we reject variable stars as references, as described in Section \ref{sec:pipeline}.

We also use the long cadence light curve to look for changes in the starspot distribution in time.
A long-term change in the observed variance due to starspots would suggest additional evidence of magnetic
cycles \citep{Mathur14, Estrela16}. 
A correlation or anti-correlation between the observed brightness of the star in the FFI photometry and 
variability from starspots would suggest that the brightness variations are dominated by faculae or
starspots, respectively \citep{Radick98}.

As a proxy for starspot variability, we calculate the \Sph\ metric from the long cadence data. 
\Sph\ measures a running standard deviation of points within five rotation periods of each cadence,
and has been shown to be a useful tracer of magnetic activity variations that requires only
photometry \citep{Mathur14}.
We then compare this photometric activity index, \Sph, to the observed bulk brightness variations 
to separate \added{stars with long-term variability dominated by spots from those with long-term 
variability dominated by facula}. \deleted{spot-dominated stars
from facula-dominated stars.}

\section{Results}

\begin{figure*}[htbp!]
\centerline{\includegraphics[width=0.9\textwidth]{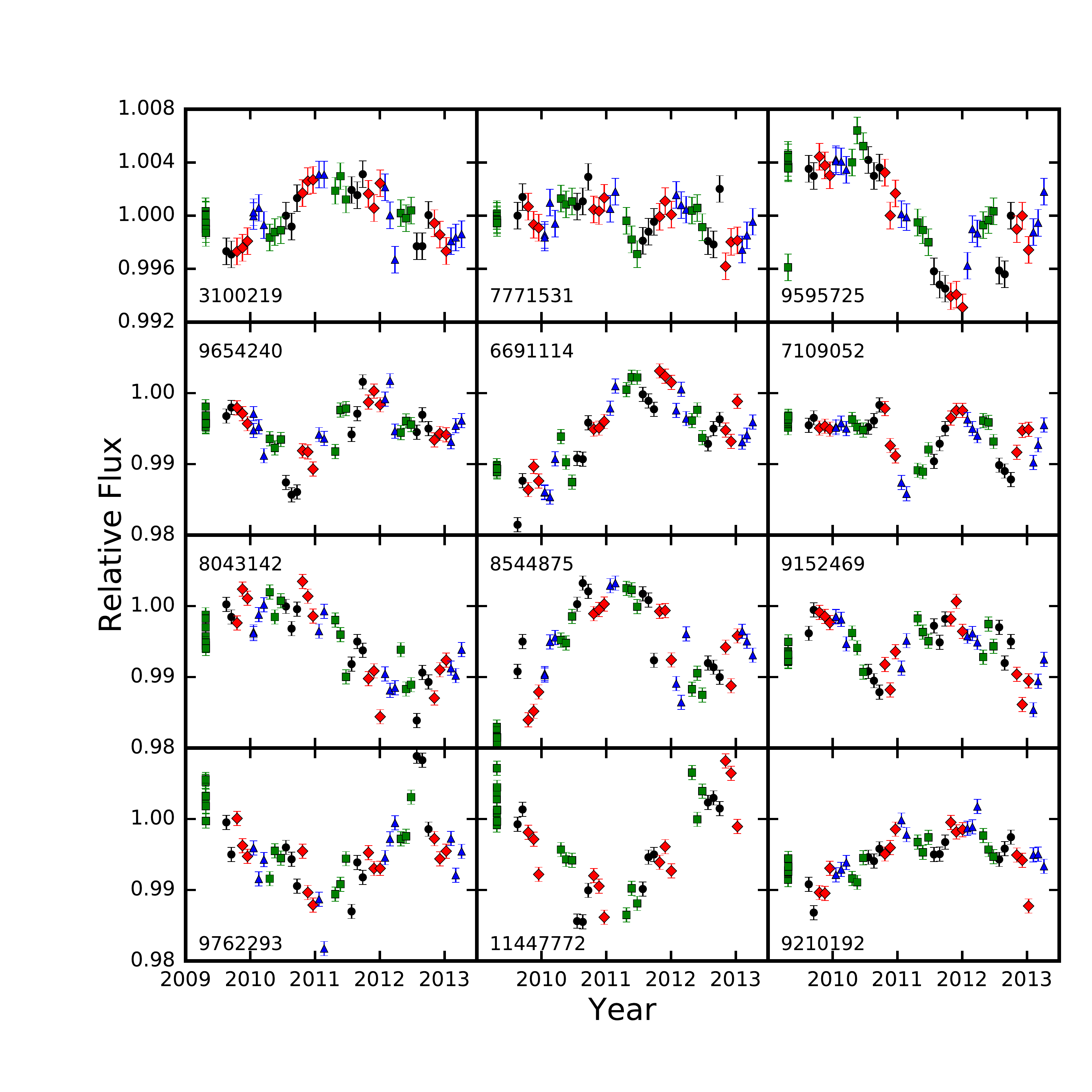}}
\caption{Representative sample of twelve stars displaying variability over the \kep\ mission. Different
stars exhibit different amplitudes of variability, with different timescales and different structure to the
observed variability. The levels of variability shown here are not observed in our control sample of 
stars with no observed rotation period, eliminating instrumental false positives.
}
\label{fig:variables}
\end{figure*}

Of the total of 3,845 stars targeted in this survey, we find that 
463 of them have observed brightness variations \added{at the $3\sigma$ level} over the \kep\
mission.
Of course, this does not mean that the remainder are unvarying: their variability is either
too small in amplitude or too low in frequency to be observed over the four-year baseline
of \kep's FFIs.
The stars with observed photometric variability are listed in Table \ref{tab:results}, and
a representative sample of observed variability is shown in Figure \ref{fig:variables}.

In almost all cases, we neglect stars with long-term trends in the photometry, only considering stars
with non-linear variability over the \kep\ mission. 
This is due to the possibility that long-term trends could be affected by data artifacts and
non-astrophysical events, as discussed in Section \ref{sec:trends}.
We claim  any other observed variability is intrinsic to the star and related to modulations in the overall
stellar brightness induced by magnetic activity.
In the following subsection, we attempt to rule out other explanations.

\begin{deluxetable*}{lcccccc}
\tablecaption{Stars with observed long-term variability during the \kep\ mission}
\footnotesize
\tablewidth{0pt}
\tablehead{
  \colhead{KIC ID} & 
  \colhead{Rotation Period\tablenotemark{a}} &
  \colhead{$K_p$\tablenotemark{b}} &
  \colhead{$B-V$ color\tablenotemark{c}} &
  \colhead{Scatter\tablenotemark{d}} &
  \colhead{Variability} &
  \colhead{Notes} \\
  \colhead{} & 
  \colhead{(days)} &
  \colhead{(mag)}  &
  \colhead{(mag)}  & 
  \colhead{(ppt)} &
  \colhead{} &
  \colhead{}      
}
\startdata
2013503 & 19.065 & 13.759 & 0.701 & 6.62 & Spots & \\
2447723 & 23.023 & 13.052 & 0.649 & 2.29 & Spots & \\
2581014 & 16.418 & 13.620 & 0.678 & 0.92 & Spots & \\
2582166 & 20.659 & 12.606 & 0.647 & 0.99 & Faculae & \\
2694675 & 12.65 & 14.335 & 0.689 & 3.77 & Spots & 3 \\
2694810 & 0.478 & 12.980 & 0.645 & 6.13 &  & 1,2,3 \\
2712799 & 8.567 & 15.138 &  0.587 & 4.89 & Spots & \\
2720200 & 10.846 & 14.882 & 0.697 & 6.83 & Spots & \\
2848223 & 4.751 & 13.620 & --- & 6.32 & Spots & \\
2974459 & 48.867 & 12.476 & 0.641 & 1.26 & Faculae & \\
\enddata
\tablenotetext{a}{Taken from \citet{McQuillan14}.}
\tablenotetext{b}{Taken from \citet{Brown11}.}
\tablenotetext{c}{Taken from \citet{Everett12}.}
\tablenotetext{d}{Standard deviation of the photometry, taking the Golden FFIs as a single point. This
estimate includes photometric uncertainty and scatter from intrinsic stellar variability together.}
\tablenotetext{1}{Short period binary}
\tablenotetext{2}{Apparent full cycle observed}
\tablenotetext{3}{GALEX-detected NUV source}
\tablecomments{Table 1 will be published in its entirety in machine-readable format on the journal website. 
A portion is reproduced here as a guide for formatting. A version is also available in the source materials
for this manuscript on the arXiv.}
\label{tab:results}
\end{deluxetable*}

\subsection{Alternative Explanations}

If the observed long-term variability were the result of an instrumental effect, then we should
expect to see similar behavior on all stars with similar stellar properties.
Fortunately, a control sample exists to test this idea. 
In addition to the stars with measured rotation periods observed by \citet{McQuillan14}, the authors
of that paper also published a list of stars without measured rotation periods. 
We select a sample of 3,000 stars that satisfy our same temperature and surface gravity cuts, but
have no measured rotation signal. 

These stars, having similar magnitudes and colors as our target sample, should provide a reasonable
control sample to compare against in a search for instrumental effects.
We repeat our analysis on this control sample and search for photometric variability. 
We find only eleven of the stars have observed variability, significantly lower than the rate
observed in the target sample.
Of these, closer inspection reveals three stars' (KIC 3338794, 4639329, and 4649300) light curves do
indeed exhibit a rotation signature in some quarters, but the signal was not detected in enough 
segments of the data to be included as a bona fide detection by \citet{McQuillan14}.
If we assume these rotation signals are real, then we have only eight stars with FFI variability but
no rotation signature, suggesting that the vast majority of our observed signals are not 
instrumental in nature.

Astrophysically, the control sample is not necessarily identical to the target sample, despite their
similar inferred temperature and gravity.
As they do not have observed rotation periods, they are likely on average to be less chromospherically
active and older. 
A fraction of them may be nearly pole-on, while others may be slightly evolved relative to their counterparts with measured rotation, leading
to their rotation signals being removed by the \kep\ data processing pipeline.

\begin{figure}[htbp!]
\centerline{\includegraphics[width=0.45\textwidth]{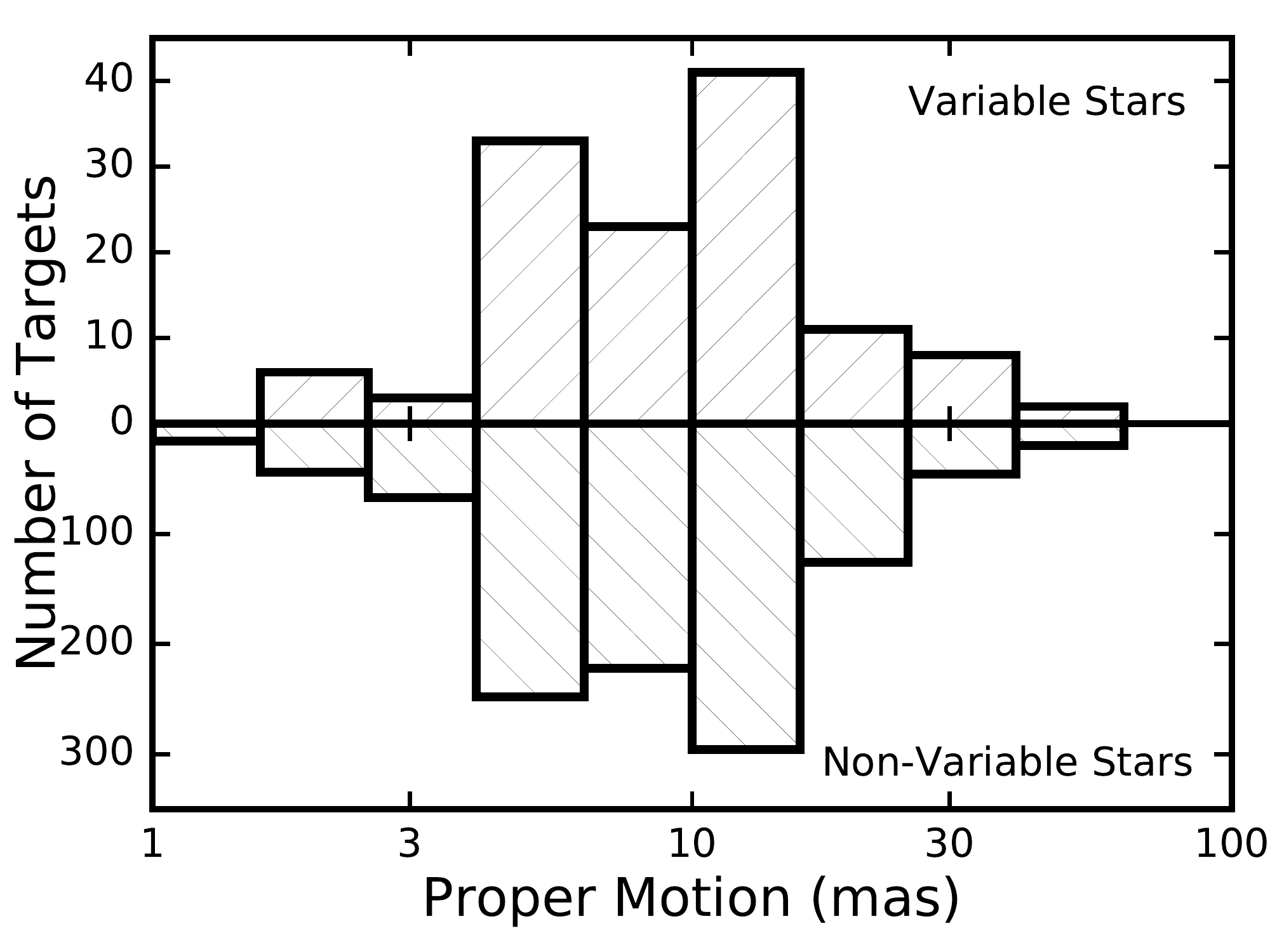}}
\caption{Proper motions of stars (top) displaying variable behavior in the FFIs and (bottom) not
displaying variable behavior in the FFIs, for all stars with listed proper motions in the Kepler
Input Catalog. The two distributions are consistent with each other, suggesting that we are not seeing
extinction from small scale structure in the ISM on preferentially high proper motion stars.
}
\label{fig:pm}
\end{figure}

\begin{figure}[htbp!]
\centerline{\includegraphics[width=0.45\textwidth]{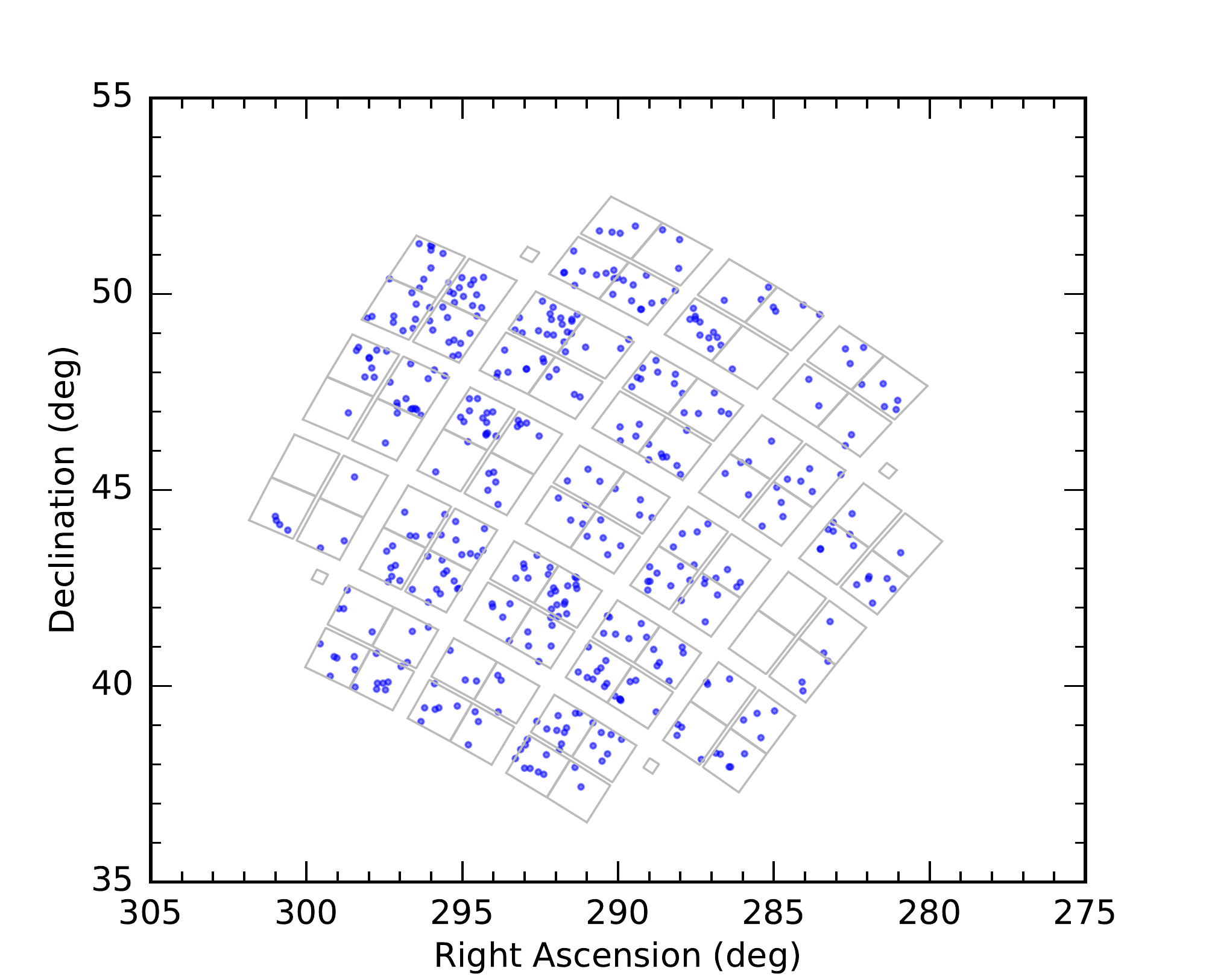}}
\caption{Targets (blue dots) with observed variability in FFI photometry. The gray lines represent
the projection of the \kep\ detector on the sky. Targets fall across the entire focal plane, roughly
matching the distribution of G stars across the detector. The galactic plane passes 8 degrees away
from the bottom left corner of the \kep\ field as shown here.
}
\label{fig:kepfield}
\end{figure}

If the stars in the control sample are systematically more distant, then they might systematically have
lower proper motions. 
Therefore, if the observed variability were caused by stars passing behind small-scale structure
in the ISM, changing the extinction along our line of sight and modulating the brightness 
\citep{Meyer96}, then
we might expect to not observe as much variability in the control sample.
This scenario is not well represented by the data either. In this case, we would expect stars with
higher proper motion to have higher rates of variability, which is not the case, as shown in Figure
\ref{fig:pm}. Moreover, we would expect stars near the galactic plane, where extinction
is higher,
to preferentially display variability, while we detect signals across the \kep\ field (Figure \ref{fig:kepfield}).

We also note that stars with observed variability are more likely to be detected in the GALEX survey of
the \kep\ field, which has a limiting NUV magnitude of 22.6 \citep{Olmedo15}.
Of the 463 target stars, 145 of them (31.3\%) are detected in the UV. Of the other 3,382 stars in our
sample, 775 of them (22.9\%) are detected in the UV, a $3.6\sigma$ discrepancy. This is in line with
expectations, as UV-bright stars are more likely to be chromospherically active \citep{Findeisen11}.

These tests rule out certain alternate explanations for the observed long-term variability, but do 
not prove that the brightness variations we are observing are magnetic in nature. The strongest evidence
in favor of that explanation is through comparing the observed variability to other tracers of
magnetic activity, such as starspots.

\subsection{Photometry as tracer of magnetic activity}

\kep\ long cadence photometry can provide information on the evolution of starspots, a proxy of
magnetic activity. 
Changes in the level of starspot variability have been shown to correlate with spectroscopic proxies of
stellar activity \citep{Mathur14, Salabert16b}. 
Starspot variability has been shown to correlate with long-term brightness variations \added{as detected through
the \Sph\ index;}
recently, C. Karoff et al. (submitted) have shown that brightness variations observed through 
\kep\ FFI photometry also correlate with 
spectroscopic and photometric proxies of magnetic activity for KIC 8006161.
\added{The same effect is seen for the Sun, where an increase in stellar chromospheric activity is visible
through both an increase in the number of observed starspots and an overall increase in flux due to the
presence of bright faculae on the surface of the star. \citep[e.g.][]{Frohlich98}.}

Here, we do not have uniform spectroscopy for our target stars, but we do have uniform long cadence 
photometry: as these stars are selected from their rotation properties, they all by definition are
on silicon during the \kep\ mission.
We compare the FFI photometry to the \Sph\ index for each of our stars with non-constant flux over
the mission.
We find that 84\% of stars exhibit behavior in their long-term
brightness that corresponds to the variability observed in the \Sph\ index. 
We label each star's brightness fluctuations as either correlated or anti-correlated in time with 
the \Sph\ index, \added{meaning the long-term variability is driven by bright faculae or dark sunspots, respectively.}
\deleted{In the latter case, we typically observe brightness variations correlated with the \Sph\ index of the star,
with the exception of a single time period with a large spot or group of spots dimming the star and 
increasing \Sph\ over part of the mission.}
The majority (68\%) are anticorrelated: increases in spot activity correspond to an overall decrease
in brightness.
This is the opposite as observed in the Sun \citep{Noyes84}. 

\subsection{A Transition in the Stellar Dynamo}
\label{sec:dynamo}

While the majority of stars have brightness variations anticorrelated with their starspots, this is
not true at all rotation periods.
This is the case because the majority of stars in our sample have rotation periods considerably shorter than
the Sun's.
Fig. 8 shows this effect. 
At rotation periods between 10 and 15 days, 11 percent of stars show detectability over the four
years of \kep\ observations; spot-dominated, anticorrelated variability is an order of magnitude more
common here than facula-dominated, correlated variability.
However, at a rotation period of 15 days, the occurrence of spot-dominated variability sharply decreases
and facula-dominated variability rises, \added{consistent with a change in the driver of the long-term brightness
variations from spots to faculae at a rotation period of approximately 24 days.}
\deleted{At a rotation period of 24 days, the occurrence of the two are equal.} We do not detect any
stars with spot-dominated variability with rotation periods beyond 26 days, while detectable
facula-dominated
variability continues to be increasingly common at rotation periods slower than the Sun.
\added{The distribution of spot-dominated stars peaks at a rotation period of 13 days; it has a mean
of 12.8 days and standard deviation of 5.6 days. The distribution of facula-dominated stars is 
visually distinct: it has a mode of 30 days, a mean of 27.5 days and a standard deviation of 9.4 days.}

We can convert observed stellar parameters into Rossby numbers ($R_0$), following the prescription of 
\citet{Noyes84} who relate a star's $B-V$ color to its convective turnover time. 
We use the $B-V$ colors of \citet{Everett12}, who performed a UBV survey of the \kep\ field largely
complete to $V \approx 19$. The typical uncertainty in each bandpass is 0.02 mag, leading to an 
uncertainty of 0.03 mag in the color and therefore $\sim 2$ days in the convective turnover time.
We see the same result in this space: stars with lower values of $R_0$ are considerably more likely
to have spot-dominated variability, while no stars with $R_0 > 1.7$ are spot-dominated.
For reference, the Sun has $R_0 = 2.05$ based on its color and rotation period.

\added{The relation between spots and faculae as the primary cause of observed long-term
photometric variability is more strongly observed in Figure 8 as a function of stellar rotation 
period than Rossby number. This effect may be due to noise in the calculation of the Rossby number.
The distribution of $B-V$ colors for our sample has a standard deviation of 0.05 mag, comparable 
in magnitude to the typical uncertainty of 0.03 mag on the $B-V$ color for any individual star. 
Additionally, a $0.03$ mag uncertainty on $B-V$ leads to a $50\%$ uncertainty on the 
convective turnover time, and thus the Rossby number. At the limit of our photometric precision,
given that our uncertainties on $B-V$ are compatible to the spread of $B-V$ colors in our sample,
Rossby number is effectively a noisy estimator of stellar rotation period, meaning rotation period
is a more suitable parameter to look for fundamental changes in stellar activity of this particular 
narrowly-defined sample.} 

Rotation period correlates with stellar age \citep{Barnes07}. \citet{Mamajek08} have shown that 
for main-sequence stars, the two are related such that
\begin{equation}
P = A^n \times a(B-V-c)^b,
\end{equation}
where $B-V$ is the color of the star in the Johnson-Cousins filters, $A$ the star's age in Myr, $P$ the
rotation period of the star in days, and $n$, $a$, $b$, and $c$ constants such that
$n = 0.566 \pm 0.008$, $a = 0.407 \pm 0.021$, $b=0.325 \pm 0.024$, and $c = 0.495\pm 0.010$. For the Sun,
$B-V = 0.653 \pm 0.003$ \citep{Ramirez12}. The stars in our sample have $B-V$ values approximately normally distributed with mean of 0.64 and standard deviation 0.04. In this case, Equation 1 reduces to
\begin{equation}
P = (0.229 \pm 0.021)\times A^{0.566 \pm 0.008}
\end{equation}

Therefore, \added{assuming the gyrochronological relations are accurate across the entire span of observed
rotation periods,} the peak occurrence of detectable spot-dominated variability occurs at an age of approximately
1200 Myr, and begins to fall off significantly at an age of 1600 Myr. \deleted{The two types of variability are
then roughly equal at an age of 3500 Myr,} \added{Stars with an age of 3500 Myr are equally likely to exhibit
long-term behavior dominated by either spots or facula,} and we do not detect any stars with spot-dominated variability
older than 4200 Myr.

\begin{figure*}[!tbp]
  \centering
  \begin{minipage}[b]{0.48\textwidth}
    \includegraphics[width=\textwidth]{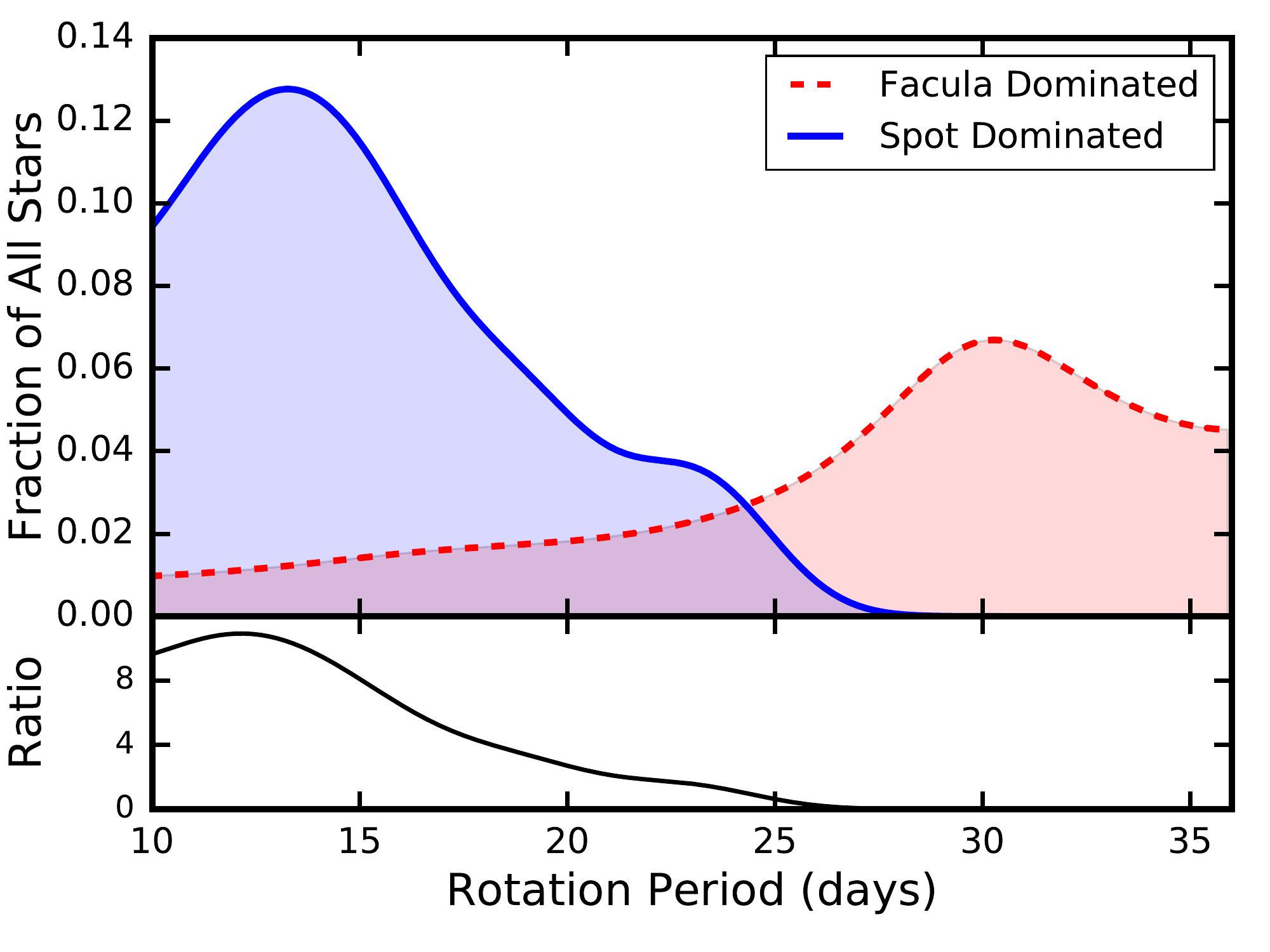}
  \end{minipage}
  \hfill
  \begin{minipage}[b]{0.48\textwidth}
    \includegraphics[width=\textwidth]{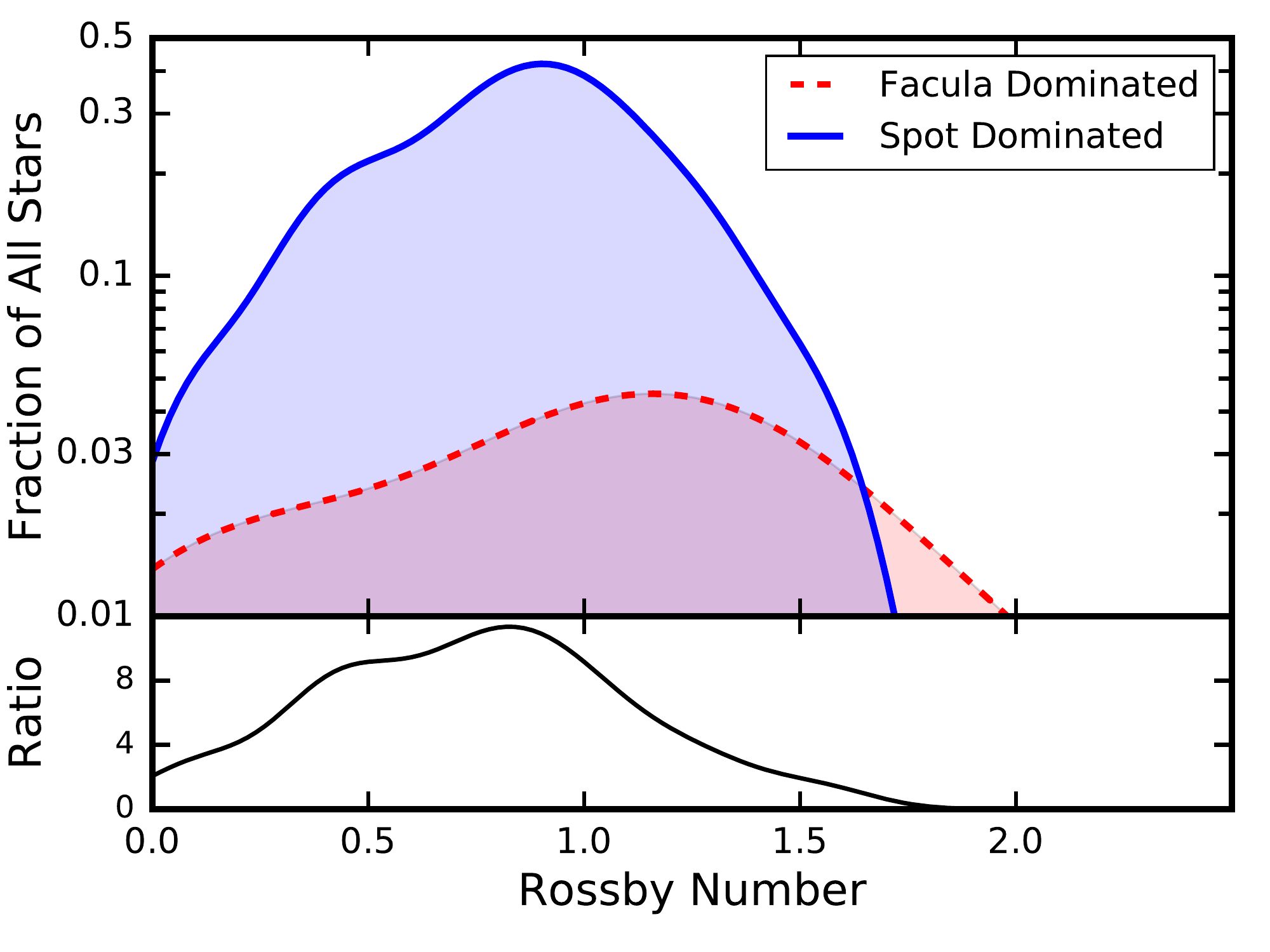}
  \end{minipage}
  \caption{(Left) Stars with observed long-term variability shown to be correlated (red) or anticorrelated (blue)
with the short-term starspot variability observed in \kep\ long cadence data.
$13\%$ of all Sun-like stars with rotation periods of 13 days have detectable spot-dominated variability.
The fraction of stars with facula-dominated variability increases in time, with the two equal at a rotation period of 24 days. Beyond 26 days, we do not detect any stars with spot-dominated variability. 
\added{These results suggest that the long-term photometric variability is driven by starspots for younger stars
and faculae for older stars, and that the transition appears to occur at a similar age for most Sun-like stars.}
(Right) The same, with Rossby number rather than rotation period on the abscissa. More than 40\% of stars
with $R_0 = 0.9$ exhibit spot-dominated variability. No stars with $R_0 > 1.7$ display spot-dominated signals: all stars wither larger Rossby numbers have facula-dominated signals. The occurrence drops off 
due to the limited time baseline and photometric precision from \kep.}
\end{figure*}

\section{Discussion}

\subsection{Transition or two populations?}

Understanding this putative transition means understanding the stars in our sample.
One help would be additional observations: it is possible that 13\% of stars with a rotation period of 13
days have observable spot-dominated
variability, or it is possible that all such stars have a signal, but it is only detectable over a 
four-year time baseline when viewed from a random angle 13\% of the time. Continued photometric
observations at this cadence and precision would be helpful, and will be obtained as described in 
Section \ref{sec:future}.

Upcoming parallax observations from Gaia will also be important for characterizing these systems. The
photometric-based effective temperature estimates for our stars have uncertainties similar in magnitude
to the range of temperatures chosen, suggesting our Sun-like stars span the F7 to G4 spectral classes.
This cut will affect different rotation periods differently. 
\added{Although there is not a large spectral range in our sample, there is a significant difference between
the typical rotation period of an average main sequence F star and G star \citep{Nielsen13}. 
Therefore, when dividing our initial sample by rotation period, the group of stars with faster rotation periods
are likely slightly biased in favor of F stars and vice versa.}
The sample of stars with rotation periods of $\sim 10$ days likely includes a higher fraction of F stars,
which slow down more slowly and have a shorter main sequence lifetime; the sample at a rotation period of
$\sim 30$ days will have a larger fraction of mid-G stars. 
As stated in the previous section, we do not detect any stars with spot-dominated variability
older than 4200 Myr; this sample by definition cannot contain stars with main sequence lifetimes shorter
than 4200 Myr.
A more detailed characterization of individual systems and the evolution of the stellar dynamo of 
G2 stars in particular will require improved stellar parameters, which will become more attainable with
measured parallaxes to these stars.

In this work, we take the gyrochronological relations at face value in our analysis that the transition
between spot-dominated and facula-dominated variability occurs at an age younger than the Sun.
Comparisons to solar twins at similar ages to the Sun show the Sun's rotation period to be
typical \citep{dosSantos16}.
Recently, \citet{VanSaders16} have shown older stars can have significantly weakened magnetic braking
compared to younger stars, leading to a discrepancy between rotation periods and age estimates,
especially for stars more evolved than the Sun.

As the vast majority of stars in our sample have ages younger than the Sun, and the transition of
interest occurs at rotation periods faster than the Sun, it is likely our results are not
significantly affected by this discrepancy, although it is certainly possible that some of 
our stars are affected.  \added{Therefore, the age estimates derived through gyrochronology in this and the 
previous 
subsection should be treated with some caution subject to the resolution of this discrepancy.}
Where possible, asteroseismic ages for stars in our sample with 
observed variability would be useful. 
As the \kep\ telescope is no longer able to point at the \kep\ field and these measurements require short
cadence data, additional data is likely needed for the vast majority of our sample in order for
such an analysis to come to fruition.
In the meantime, spectroscopic activity indicators would be useful to probe the putative transition and
compare to the spectroscopic transition considered by previous studies, such as that of \citet{Metcalfe16}.

\subsection{Comparison with previous work}
\citet{Metcalfe16} have suggested spectroscopic diagnostics indicate that the solar dynamo may be in
transition, due to the loss of the large-scale stellar magnetic field.
Although we find a possible transition at a younger age, around 3 Gyr, these two results are not necessarily
inconsistent.
The transition described in that paper, which matches the theoretical work of \citet{VanSaders16},
would suggest that weakened magnetic braking in older stars would lead to a pileup of rotation periods
as a function of spectral type. 
If the collection of spot-dominated stars we observe with rotation periods of $\approx 15$ days were
mostly late F stars at a variety of ages, and the facula-dominated stars with rotation periods of $\approx$ 25
days were mostly G stars, then this would support these predictions.
As before, improved stellar properties and ages would be useful to better understand the relation between
the potential transition discussed here and the spectroscopic transition discussed in that paper.

\citet{Olah16} found that there is a transition between smooth and complex magnetic cycles at an age
of 2 to 3 Gyr of age. This is in line with our analysis, as we find a rapid decrease in the fraction
of stars with spot dominated variability during this age rage, which corresponds to rotation periods
of 17-22 days. It is plausible that the more rapidly rotating, spot-dominated stars contain short-period and
long-period
cycles \citep[e.g][]{Brandenburg17}, leading to a more complex appearance in ground-based data. 
Additional long-term observations of these stars, leading to a detection of longer-timescale magnetic cycles
in these same stars, would support this claim.

\citet{Saar99} and \citet{Bohm-Vitense07} divide stars with observed cycles into ``active'' and ``inactive''
branches. Some of these stars have cycles that would be observable over the relatively short time baseline
of \kep. These shorter cycle periods are limited by observational biases from the ground and are not
well-sampled. Data from \kep\ provide an opportunity to identify systems to compare to the ground-based
detections of magnetic cycles along both of these branches.

\subsection{Particularly Interesting Systems}

\subsubsection{Possibly Complete Magnetic Cycles}

A four-year time baseline does not preclude the detection of magnetic cycles.
Previous studies have shown the existence of magnetic cycles lasting 2-3 years on stars \citep[e.g.][]{Saar99}.
We identify by eye 28 systems which appear to exhibit at least one cycle over the baseline of \kep. 
\citet{Brandenburg17} have shown stars with longer cycles can have additional, shorter cycles
with periods of 1-3 years.
Therefore, it is not implausible that some of our stars may have considerably longer cycles in 
addition to the ones observed in \kep\ FFI photometry which will reveal themselves through continued
observations.

\begin{figure}[htbp!]
\centerline{\includegraphics[width=0.45\textwidth]{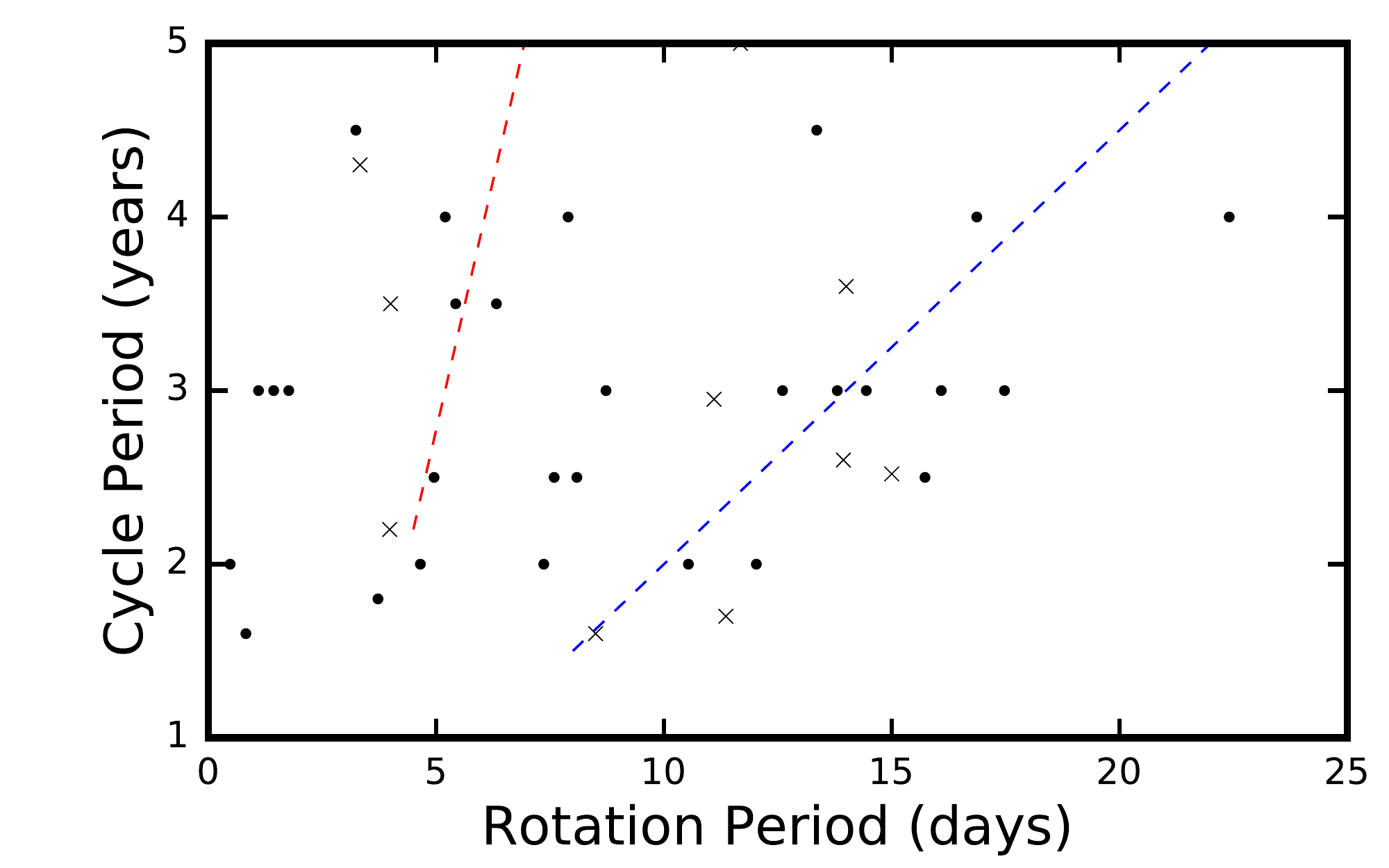}}
\caption{Rotation periods and magnetic cycle periods for stars with apparently complete 
observed magnetic cycles (black points). Red and blue dashed lines represent the inactive 
and active branches of stellar activity as defined by \citet{Bohm-Vitense07}. Points marked 
with an 'x' are values from the literature, taken from \citet{Saar99}, 
\citet{Metcalfe10, Metcalfe13}, and \citet{Egeland15}. While there is scatter, \added{the observed
points are not consistent with being randomly placed in rotation period-cycle length space}.
}
\label{fig:cycleperiods}
\end{figure}

The observed cycles and rotation periods for these stars are plotted in Fig. \ref{fig:cycleperiods},
as well as similar stars in the literature. We see that many of these stars follow the ``active''
and ``inactive'' branches of \citet{Bohm-Vitense07}. 
\added{We estimate the scatter by calculating the RMS distance from each of the 38 points to the closer 
of
the two lines representing the active and inactive branches on this figure, and compare
it against the RMS distance from 38 randomly drawn points in the same plane. 
We draw rotation periods following the distribution of rotation periods in our sample,
and draw activity cycle periods uniformly over the range shown in Fig. 9.
}

\added{The distribution of randomly drawn points is significantly further away from the
two branches than the observed potential magnetic cycles. 
In 500,000 simulated data sets, only 24 of them (0.0048\%) have an RMS scatter smaller than
the real data set.
If we only consider the 28 points in our sample on this figure, ignoring the 10 points
from previous analyses, then 260 of 500,000 simulated data sets have RMS scatter smaller
than the real data set. 
Both of these are significant at the $3\sigma$ level.
The difference between these two subsamples is due to the increased variance in the random
distribution when selecting a smaller number of test samples: the difference in scatter between
our own targets and the two branches compared to the stars from previous studies is not 
significant.
}

Additional photometry of these stars to confirm
these cycles, \added{better measure their periods,} and search for longer cycles that are also
possibly present in these stars can provide
constraints for future models of stellar dynamos. 
In Table 2 we list those systems observed to have apparently complete magnetic cycles, with an
estimate of the observed cycle period. Two of these stars, each with very short periods and likely
binary systems, have their cycles noted in the \Sph\ index but not in the FFI data, suggesting the cycle
may be on the fainter component which does not dominate the FFI photometry.
These two stars are noted in the Table.
Additional observations of these systems could help understand the evolution of magnetic cycles of binary 
systems.

\begin{deluxetable}{lcc}
\tablecaption{Stars with apparently complete magnetic cycles}
\footnotesize
\tablewidth{0pt}
\tablehead{
  \colhead{KIC ID} & 
  \colhead{Semiamplitude} &
  \colhead{Period} \\
  \colhead{} & 
  \colhead{(percent)} &
  \colhead{(yr)} 
}
\startdata
2694810 & 0.6 & 2 \\
3236788 & 2.0 & 3.5 \\
3743810 & 0.3 & 3 \\
4555566 & 1.2 & 2 \\
4726114 & 0.3 & 4 \\
5352687 & 0.9 & 4 \\
5450764 & 0.7 & 4 \\
6038355 & 0.8 & 2.5 \\
6263983 & 1.0 & 3 \\
6708110 & 1.2 & 3 \\
7272437 & 0.1 & 3 \\
7432092 & 0.3 & 4 \\
7433192 & 0.2 & 3 \\
7678238 & 0.6 & 2.5 \\
8041424 & 1.8 & 1.8 \\
8043142 & 0.5 & 4.5 \\
8345997 & 0.4 & 3 \\
8759594 & 0.2 & 2.5 \\
8804069 & 0.9 & 3 \\
9306271\tablenotemark{1} & --- & 1.6 \\
10087863 & 0.6 & 3 \\
10122937 & 0.9 & 3.5 \\
10921242\tablenotemark{1} & --- & 3 \\
11014223 & 1.5 & 2 \\
11033434 & 0.6 & 2 \\
11415049 & 0.3 & 2 \\
11873617 & 0.7 & 2.5 \\
12417799 & 1.2 & 4.5 \\
\enddata
\tablenotetext{1}{Cycle observed and period inferred from \Sph\ variability}
\label{tab:cycles}
\end{deluxetable}

\subsubsection{Short-Period Binaries}
\label{sec:cb}
Several of the stars observed with the most extreme long-term brightness variations have, on inspection
of their long cadence light curves, been revealed as short-period eclipsing binaries.
It has long been considered that contact binaries could have extreme magnetic activity which leads to 
an evolution of their brightness \citep{Applegate92}. 
Such brightness variations have been detected, as has indirect evidence for short period 
magnetic cycles on contact biaries \citep{Ibanoglu01, Borkovits05}.

Recently, \citet{Marsh17} analyzed nearly 10,000 contact binaries observed in the 
Catalina Real-Time Transient Survey \citep{Djorgovski11}, finding that 20\% of them undergo a 
linear change in brightness during the survey. 

\begin{figure}[htbp!]
\centerline{\includegraphics[width=0.45\textwidth]{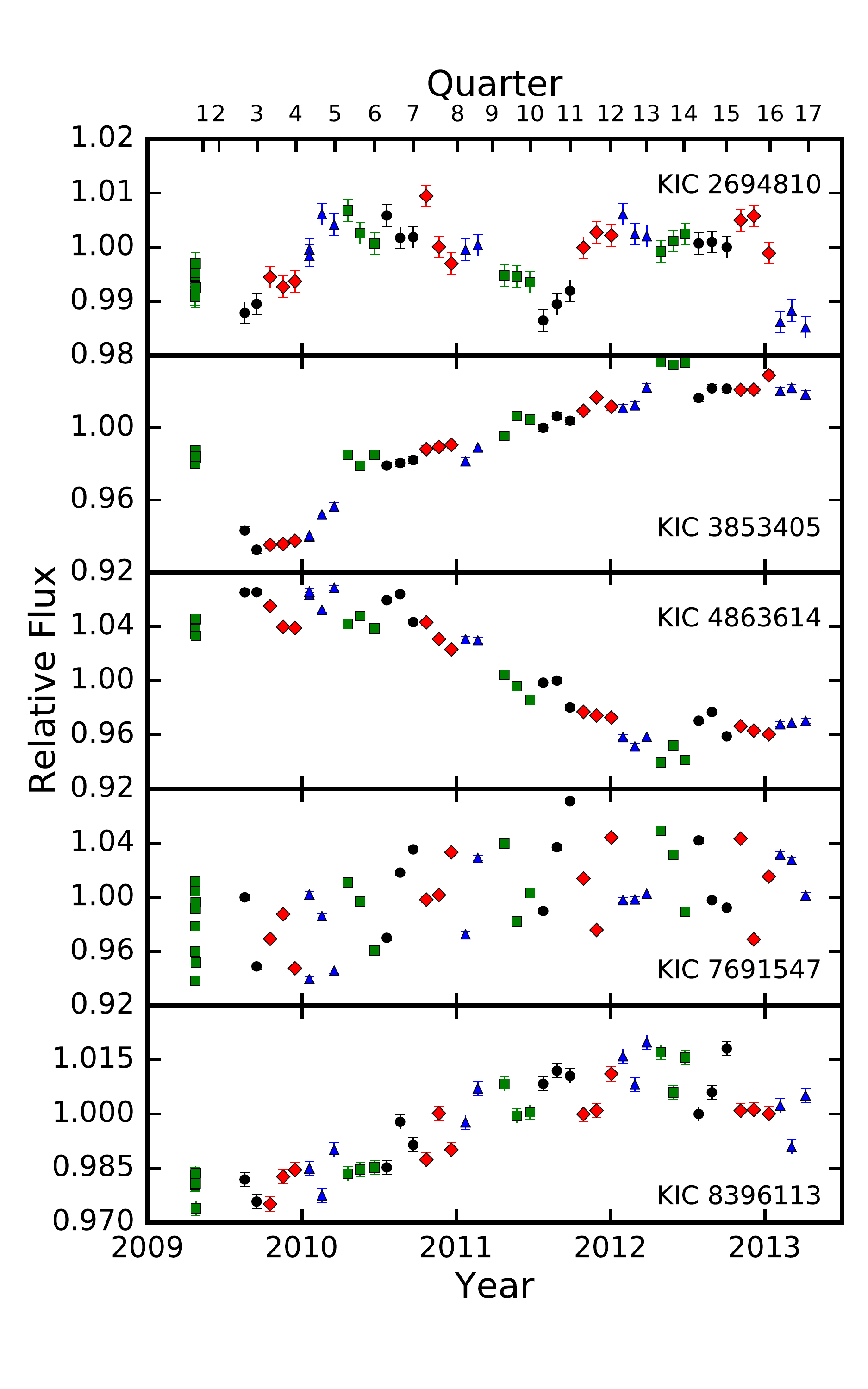}}
\caption{\textit{f3} time series photometry from five short-period binaries observed in this work displaying
long-term
variability which may be evidence for extreme magnetic cycles in these stars.
Uncertainties correspond only to the photometric uncertainties on the individual observations, not to
intrinsic stellar variability on the host star itself. For example, KIC 7691547 displays 4\% variability
peak to peak during each 8-hour orbit, which can be seen as excess scatter in the FFI data but is 
not accounted for in the uncertainties. Combining long cadence photometry with FFI data could be used for
detailed modeling of the evolution of these systems.
}
\label{fig:binaries}
\end{figure}

Here, we present an additional five short-period binaries with long-term FFI variability.
A detailed analysis of their evolution is beyond the scope of this work, but we present their
light curves in Fig. \ref{fig:binaries} as a potential avenue for further follow-up observations.
In general, the detailed long cadence light curves from \kep\ enable the opportunity to characterize
both the evolution in the eclipse-to-eclipse variability over the mission in concert with long-term
photometric variability. 

While the changes in brightness for contact binaries have been traditionally considered to be due to
magnetic effects, other effects can cause long-term photometric trends in the observed photometry as well.

\subsection{Long-Term Photometric Trends}
\label{sec:trends}

In this work, we largely ignore systems with long-term trends observed in the \kep\ FFI photometry.
While we can eliminate most false positives for other stars with variability, meaning the results in
Table 1 should be dominated by stars with magnetic activity variations, some false positives remain for
systems with long-term trends.
One false positive that may be common is faint stars in the field of view.
The typical \kep\ PSF has a radius of 6 arcsec \citep{Bryson10}; the Kepler Input Catalog only
includes targets down to 21st magnitude and is incomplete at the faint end \citep{Brown11}.

A faint M dwarf with a high proper motion could then be found to be expanding its PSF into the aperture
for any target, leading to a long-term trend in the data.
With the current data, we are unable to test this. 
Detailed PSF modeling of each star may be able to improve this in the future: if certain stars have 
changing levels of contamination, then their PSF should be slightly varying relative to other nearby 
stars.

In some cases, long-term trends can also be caused by varying levels of extinction. This may be especially
true for contact binaries or stars near the end of their life cycles, where shells of ejected
material are expanding and decreasing in opacity in time.
KIC 3853405 is an example of a system that displays this behavior.
This contact binary increases in brightness by more than 5\% over the \kep\ mission 
(Fig. \ref{fig:binaries}). 
While described in the previous subsection as potential magnetic variability,
the variation could also be explained by a
change in the opacity of circumstellar dust ejected in a recent outburst
or by an actual increase in luminosity as the two stellar cores prepare to merge \citep{Tylenda11, Molnar17}.
Detailed modeling of both the long cadence light curves and FFI photometry, combined with additional
follow-up observations could be used to better understand this system, but is beyond the scope of this
paper.

In Table 1 and throughout this work, we reject nearly all systems with long-term trends as they
cannot be uniquely shown to be related to changes of the magnetic activity of the stars.
The only exceptions are stars in which the \Sph\ index is also changing in a linear way throughout the
\kep\ mission.
As the apertures used in long cadence photometry are substantially smaller than ours, we do not 
necessarily expect FFI brightness variations from stellar contamination to dilute the long cadence
light curves in the same way, so we argue these changes are likely to be astrophysical. We specifically
note these systems in Table 1, and also note their inclusion or exclusion does not significantly 
affect the results of our analysis.

\section{Conclusions}

\subsection{FFIs as a probe of long-term astrophysical variability}

In this paper, we have presented the method behind the \textit{f3} package to produce very long cadence
light curves from \kep\ FFI data. 
These light curves provide photometry at monthly cadence at 52 epochs across the \kep\ mission,
and generally avoid the removal of astrophysical variations on month-long or slower timescales inherent
in the \kep\ long cadence data.

Accurate relative photometry on long time scales is important, as it can be used to probe long-term
behavior of the brightness of astrophysical objects typically removed in transit surveys like
\kep.
Here we develop light curves for a sample of Sun-like stars with observed rotation periods to probe
changes in stellar activity in time, finding that we do observe evidence for magnetic activity in a
substantial fraction of Sun-like stars, especially those that are young and rapidly rotating.
We find that there appears to be a transition in the stellar dynamo at a stellar rotation period
of 15-25 days, corresponding to a gyrochronological age of 1.5-3.5 Gyr, although this may reflect 
a change in stellar parameters, with younger, hotter stars in our sample more likely to be rapidly rotating
than older, cooler stars.
Stars with rotation periods faster than 15 days have photometric variations that are typically dominated by
starspots, while more slowly rotating stars have variations dominated by faculae.
Additionally, we find that the more rapidly rotating stars are typically significantly more active 
than the slowly rotating stars, in line with previous work.
We also identify 28 stars with apparently complete short-period magnetic cycles, finding they are
consistent with the ``active'' and ``inactive'' branches of stars of \citet{Saar99}.

Although it is our application in this work, this method is not exclusive to observing magnetic cycles.
In principle, photometry from \textit{f3} could be used to probe intrinsic brightness variations from 
variable stars, variable extinction through the ISM, or evolution of binary stars or evolved stars
which affects their observed brightness.
To explore this dataset to the fullest extent possible, we make our code publicly available for community
use and development.

\subsection{Future prospects}
\label{sec:future}

Future missions will enable similar measurements on even larger samples of stars.
\TESS\ \citep{Ricker14} will obtain a FFI every $\sim 20$ minutes, enabling aperture 
photometry with a large number
of reference stars at a high cadence. 
However, each field will only be observed for one month at a time, and the continuous viewing zone
will only span one year of observations: 
longer time baselines will only be possible with an extended mission. 
Stars in the continuous viewing zone will fall on 13 different pixels for one month each during
the mission, so long-term changes like those observed in this field will require detailed
knowledge of the instrument's flat field. 
We encourage the \TESS\ team to make every effort to understand the flat field both before launch
and during the mission.

Another possibility to detect magnetic cycles on a large number of stars is provided by the \Gaia\ mission
\citep{Prusti16}. 
While its photometric performance is often overlooked relative to its astrometric potential,
\Gaia\ will return $\approx 1$ mmag photometry for all stars with $G < 14$ and $\approx 2$ mmag photometry
for all stars with $G < 16$ \citep{Jordi10}. 
With the typical star observed 70 times over 10 years, \Gaia\ will provide
FFI-like photometry at FFI-like cadence for millions of stars for a decade. 
\Gaia\ long-term photometry combined with rotation periods from simultaneous observations with
\kt, \TESS, and possibly \textit{PLATO} \citep{Rauer14} will enable detailed photometric studies of 
magnetic cycles for millions of stars across the galaxy, allowing us to understand the interplay between
magnetic activity and photometric modulations across the H-R diagram.

\acknowledgements

We thank Geert Barentsen for conversations that improved the quality of the figures in this
manuscript. We also thank Eric Ford, Jacob Bean, Travis Metcalfe, and 
Rachael Roettenbacher for helpful discussions in the 
preparation of this manuscript, and Megan Bedell for naming the \textit{f3} package.

Work by B.T.M. and D.F.-M. was performed under contract with the Jet Propulsion
Laboratory (JPL) funded by NASA through the Sagan Fellowship Program executed
by the NASA Exoplanet Science Institute.
G.T. acknowledges support from the Banneker Institute at the Harvard-Smithsonian
Center for Astrophysics.

This paper includes data collected by the \kep\ mission. Funding for the
\kep\ mission is provided by the NASA Science Mission directorate.
We are grateful to the entire \kep\ team, past and present.
These data were obtained from the Mikulski Archive for Space Telescopes
(MAST).
STScI is operated by the Association of Universities for Research in
Astronomy, Inc., under NASA contract NAS5-26555.
Support for MAST is provided by the NASA Office of Space Science via grant
NNX13AC07G and by other grants and contracts.

\software{%
    numpy \citep{numpy},
    matplotlib \citep{matplotlib},
    Theano \citep{Theano},
    mahotas \citep{mahotas},
    K2fov \citep{Mullally16},
    f3}

\facility{Kepler}

\appendix

\section{Probabilistic calibration model}\label{sect:probcal}

We model the relative flux $f_{n,t}$ of star $n$ at time $t$ as
\begin{eqnarray}
\hat{f}_{n,t} = r_{n,m(t)}\,z_{t}
\end{eqnarray}
where the hat indicates this is the predicted model value of $f_{n,t}$,
$r_{n,s(t)}$ is the total response of the pixels in the aperture for the star
at the season $m(t)$ of exposure $t$, and $z_t$ is the time-variable mean zero
point of exposure $t$.
\added{Since we use aperture photometry to measure the flux with different apertures
on different detectors in each season, the relative photometric calibration
between seasons is not known \emph{a priori}.
We capture this effect by fitting for the parameters $r_{n,s(t)}$, one for each
target in each season.
Then, for this model, we assume that the mean zero-point for a set of nearby
targets varies systematically and we capture this variability by fitting for
a parameter $z_t$ that is shared by all targets at a single exposure $t$.
Unlike the \kep\ long- and short-cadence light curves, the dominant source
of systematic variability on the timescales relevant for the FFIs is not
pointing variation.
Instead, it is longer timescale trends---like temperature variations and
detector degration---that affect targets similarly.}

\added{To complete our probabilistic model, we must also specify a noise model
for each measurement.
As ususal, there is a contribution from the intrinsic measurement uncertainty
$\sigma$, but that is not the dominant noise source.
We must also take the intrinsic variability of the star and any variance
introduced by misspecification of the calibration model into account.}
To capture these effects, we model the variance for each observation as
\begin{eqnarray}
{\sigma_{n,t}}^2 = {V_n}^2\,{\hat{f}_{n,t}}^2 + {S_t}^2\,{r_{n,s(t)}}^2 +
    \sigma^2
\end{eqnarray}
where $V_n$ is the amplitude of the intrinsic variability of star $n$,
$S_t$ is the scatter in the zero point of the exposure $t$, and $\sigma$
is photometric noise level.
\added{In this model, we simultaneously fit for the parameters $V_n$, $S_t$,
and $\sigma$ along with the calibration parameters.
This model can be interpretated as a quantification of the intuition that
variable stars should contribute less weight to the calibration and that
some epochs will be intrinsically noisier.}
To break the degeneracy between the pixel responses and the zero points, we
regularize the fit by selecting Gaussian priors with unit mean and variance
$10^{-4}$ for both $r_{n,m(t)}$ and $z_t$.

The model described in the previous paragraph assumes that the measurements in
the light curve are independent.
Since the observations are not uniformly distributed in time, this can lead to
biased estimates of the pixel responses for targets with long-term
variability---exactly the case that we are interested in.
To mitigate this issue, we model the true flux of the target star using a
Gaussian Process with a Mat\'ern-3/2 kernel.
This implies a log-likelihood function for the target star of
\begin{eqnarray}
\mathcal{L}_\mathrm{0} (\boldsymbol{\theta}) =
    -\frac{1}{2}\,(\boldsymbol{f}_0 - \boldsymbol{\hat{f}}_0)^{\mathrm{T}}
    \,K^{-1}\,(\boldsymbol{f}_0 - \boldsymbol{\hat{f}}_0)
    - \frac{1}{2}\,\log\det K - \frac{T}{2}\,\log 2\,\pi
\end{eqnarray}
where we have labeled the target star with $n=0$ and the elements of the
covariance matrix $K$ are given by
\begin{eqnarray}
K_{i,j} = \delta_{ij}\,(\sigma^2 + {S_i}^2\,{r_{0,s(i)}}^2)
    + \alpha\,\left(1+\sqrt{3}\frac{|t_i-t_j|}{\tau} \right)\,\exp\left(
    -\sqrt{3}\frac{|t_i-t_j|}{\tau}
    \right)
\end{eqnarray}
where $\delta_{ij}$ is the Kronecker delta, $T$ is the number of cadences in the
light curve, and $\alpha$ and $\tau$ are parameters of the model.

Applying this model to a target star (labeled $n=0$) and set of $N$ reference stars
observed at $T$ cadences, we obtain the log-likelihood function
\begin{eqnarray}\label{eq:likelihood}
\mathcal{L}(\boldsymbol{\theta}) &=& \mathcal{L}_0(\boldsymbol{\theta})
-\frac{1}{2}\sum_{n=1}^N\,\sum_{t=1}^T
\left[
\frac{(f_{n,t} - \hat{f}_{n,t})^2}{{\sigma_{n,t}}^2}
+ \log \left( 2\,\pi\,{\sigma_{n,t}}^2 \right)
\right] \nonumber\\
&& - \frac{1}{2}\sum_{t=1}^T\frac{(z_t - 1)^2}{10^{-4}}
- \frac{1}{2}\sum_{n=0}^N\,\sum_{m=1}^4\frac{(r_{n,m} - 1)^2}{10^{-4}}
\end{eqnarray}
where $\boldsymbol{\theta}$ refers to the set of the following parameters:
\begin{itemize}
\item $4\,(N+1)$ parameters $r_{n,m(t)}$,
\item $T$ parameters $z_t$,
\item $N$ parameters $V_n$,
\item $T$ parameters $S_t$,
\item a single noise parameter $\sigma$, and
\item the 2 hyperparameters of the Gaussian Process, $\alpha$ and $\tau$.
\end{itemize}
To fit for the maximum likelihood parameters $\boldsymbol{\theta}^*$, we build this model using
{\it Theano} \citep{Theano} and maximize Equation~(\ref{eq:likelihood}) using 2000 iterations of
the {\it Adam} algorithm \citep{Kingma14}.

Conditioned on these maximum likelihood parameters, the de-trended flux for star $n$ at time $t$ is
given by
\begin{eqnarray}
\frac{f_{n,t}}{{\hat{f}_{n,t}}*}
\end{eqnarray}
with predictive variance
\begin{eqnarray}
\frac{{{S_t}^*}^2\,{{r_{n,s(t)}}^*}^2 + {\sigma^*}^2}{{{\hat{f}_{n,t}}*}^2} \quad.
\end{eqnarray}


\begin{thebibliography}{}
\expandafter\ifx\csname natexlab\endcsname\relax\def\natexlab#1{#1}\fi
\providecommand{\url}[1]{\href{#1}{#1}}

\bibitem[{{Angus} {et~al.}(2015){Angus}, {Aigrain}, {Foreman-Mackey}, \&
  {McQuillan}}]{Angus15}
{Angus}, R., {Aigrain}, S., {Foreman-Mackey}, D., \& {McQuillan}, A. 2015,
  \mnras, 450, 1787

\bibitem[{{Applegate}(1992)}]{Applegate92}
{Applegate}, J.~H. 1992, \apj, 385, 621

\bibitem[{{Baliunas} {et~al.}(1995){Baliunas}, {Donahue}, {Soon}, {Horne},
  {Frazer}, {Woodard-Eklund}, {Bradford}, {Rao}, {Wilson}, {Zhang}, {Bennett},
  {Briggs}, {Carroll}, {Duncan}, {Figueroa}, {Lanning}, {Misch}, {Mueller},
  {Noyes}, {Poppe}, {Porter}, {Robinson}, {Russell}, {Shelton}, {Soyumer},
  {Vaughan}, \& {Whitney}}]{Baliunas95}
{Baliunas}, S.~L., {Donahue}, R.~A., {Soon}, W.~H., {et~al.} 1995, \apj, 438,
  269

\bibitem[{{Barnes}(2007)}]{Barnes07}
{Barnes}, S.~A. 2007, \apj, 669, 1167

\bibitem[{{Bastien} {et~al.}(2016){Bastien}, {Stassun}, {Basri}, \&
  {Pepper}}]{Bastien16}
{Bastien}, F.~A., {Stassun}, K.~G., {Basri}, G., \& {Pepper}, J. 2016, \apj,
  818, 43

\bibitem[{{B{\"o}hm-Vitense}(2007)}]{Bohm-Vitense07}
{B{\"o}hm-Vitense}, E. 2007, \apj, 657, 486

\bibitem[{{Borkovits} {et~al.}(2005){Borkovits}, {Elkhateeb}, {Csizmadia},
  {Nuspl}, {B{\'{\i}}r{\'o}}, {Heged{\"u}s}, \& {Csorv{\'a}si}}]{Borkovits05}
{Borkovits}, T., {Elkhateeb}, M.~M., {Csizmadia}, S., {et~al.} 2005, \aap, 441,
  1087

\bibitem[{{Borucki} {et~al.}(2010){Borucki}, {Koch}, {Basri}, {Batalha},
  {Brown}, {Caldwell}, {Caldwell}, {Christensen-Dalsgaard}, {Cochran},
  {DeVore}, {Dunham}, {Dupree}, {Gautier}, {Geary}, {Gilliland}, {Gould},
  {Howell}, {Jenkins}, {Kondo}, {Latham}, {Marcy}, {Meibom}, {Kjeldsen},
  {Lissauer}, {Monet}, {Morrison}, {Sasselov}, {Tarter}, {Boss}, {Brownlee},
  {Owen}, {Buzasi}, {Charbonneau}, {Doyle}, {Fortney}, {Ford}, {Holman},
  {Seager}, {Steffen}, {Welsh}, {Rowe}, {Anderson}, {Buchhave}, {Ciardi},
  {Walkowicz}, {Sherry}, {Horch}, {Isaacson}, {Everett}, {Fischer}, {Torres},
  {Johnson}, {Endl}, {MacQueen}, {Bryson}, {Dotson}, {Haas}, {Kolodziejczak},
  {Van Cleve}, {Chandrasekaran}, {Twicken}, {Quintana}, {Clarke}, {Allen},
  {Li}, {Wu}, {Tenenbaum}, {Verner}, {Bruhweiler}, {Barnes}, \&
  {Prsa}}]{Borucki10}
{Borucki}, W.~J., {Koch}, D., {Basri}, G., {et~al.} 2010, Science, 327, 977

\bibitem[{{Brandenburg} {et~al.}(2017){Brandenburg}, {Mathur}, \&
  {Metcalfe}}]{Brandenburg17}
{Brandenburg}, A., {Mathur}, S., \& {Metcalfe}, T.~S. 2017, ArXiv e-prints,
  arXiv:1704.09009

\bibitem[{{Brown} {et~al.}(2011){Brown}, {Latham}, {Everett}, \&
  {Esquerdo}}]{Brown11}
{Brown}, T.~M., {Latham}, D.~W., {Everett}, M.~E., \& {Esquerdo}, G.~A. 2011,
  \aj, 142, 112

\bibitem[{{Bryson} {et~al.}(2010){Bryson}, {Tenenbaum}, {Jenkins},
  {Chandrasekaran}, {Klaus}, {Caldwell}, {Gilliland}, {Haas}, {Dotson}, {Koch},
  \& {Borucki}}]{Bryson10}
{Bryson}, S.~T., {Tenenbaum}, P., {Jenkins}, J.~M., {et~al.} 2010, \apjl, 713,
  L97

\bibitem[{{Charbonneau}(2010)}]{Charbonneau10}
{Charbonneau}, P. 2010, Living Reviews in Solar Physics, 7, 3

\bibitem[{Coelho(2013)}]{mahotas}
Coelho, L.~P. 2013, Journal of Open Research Software, 1, doi:10.5334/jors.ac.
\newblock \url{http://dx.doi.org/10.5334/jors.ac}

\bibitem[{{Davenport}(2016)}]{Davenport16}
{Davenport}, J.~R.~A. 2016, \apj, 829, 23

\bibitem[{{De Cat} {et~al.}(2015){De Cat}, {Fu}, {Ren}, {Yang}, {Shi}, {Luo},
  {Yang}, {Wang}, {Zhang}, {Shi}, {Zhang}, {Dong}, {Catanzaro}, {Corbally},
  {Frasca}, {Gray}, {Molenda-{\.Z}akowicz}, {Uytterhoeven}, {Briquet},
  {Bruntt}, {Frandsen}, {Kiss}, {Kurtz}, {Marconi}, {Niemczura}, {{\O}stensen},
  {Ripepi}, {Smalley}, {Southworth}, {Szab{\'o}}, {Telting}, {Karoff}, {Silva
  Aguirre}, {Wu}, {Hou}, {Jin}, \& {Zhou}}]{DeCat15}
{De Cat}, P., {Fu}, J.~N., {Ren}, A.~B., {et~al.} 2015, \apjs, 220, 19

\bibitem[{{Djorgovski} {et~al.}(2011){Djorgovski}, {Drake}, {Mahabal},
  {Graham}, {Donalek}, {Williams}, {Beshore}, {Larson}, {Prieto}, {Catelan},
  {Christensen}, \& {McNaught}}]{Djorgovski11}
{Djorgovski}, S.~G., {Drake}, A.~J., {Mahabal}, A.~A., {et~al.} 2011, ArXiv
  e-prints, arXiv:1102.5004

\bibitem[{{dos Santos} {et~al.}(2016){dos Santos}, {Mel{\'e}ndez}, {do
  Nascimento}, {Bedell}, {Ram{\'{\i}}rez}, {Bean}, {Asplund}, {Spina},
  {Dreizler}, {Alves-Brito}, \& {Casagrande}}]{dosSantos16}
{dos Santos}, L.~A., {Mel{\'e}ndez}, J., {do Nascimento}, J.-D., {et~al.} 2016,
  \aap, 592, A156

\bibitem[{{Egeland} {et~al.}(2015){Egeland}, {Metcalfe}, {Hall}, \&
  {Henry}}]{Egeland15}
{Egeland}, R., {Metcalfe}, T.~S., {Hall}, J.~C., \& {Henry}, G.~W. 2015, \apj,
  812, 12

\bibitem[{{Endl} {et~al.}(2016){Endl}, {Brugamyer}, {Cochran}, {MacQueen},
  {Robertson}, {Meschiari}, {Ramirez}, {Shetrone}, {Gullikson}, {Johnson},
  {Wittenmyer}, {Horner}, {Ciardi}, {Horch}, {Simon}, {Howell}, {Everett},
  {Caldwell}, \& {Castanheira}}]{Endl16}
{Endl}, M., {Brugamyer}, E.~J., {Cochran}, W.~D., {et~al.} 2016, \apj, 818, 34

\bibitem[{{Estrela} \& {Valio}(2016)}]{Estrela16}
{Estrela}, R., \& {Valio}, A. 2016, \apj, 831, 57

\bibitem[{{Everett} {et~al.}(2012){Everett}, {Howell}, \&
  {Kinemuchi}}]{Everett12}
{Everett}, M.~E., {Howell}, S.~B., \& {Kinemuchi}, K. 2012, \pasp, 124, 316

\bibitem[{{Findeisen} {et~al.}(2011){Findeisen}, {Hillenbrand}, \&
  {Soderblom}}]{Findeisen11}
{Findeisen}, K., {Hillenbrand}, L., \& {Soderblom}, D. 2011, \aj, 142, 23

\bibitem[{{Fossati} {et~al.}(2017){Fossati}, {Marcelja}, {Staab}, {Cubillos},
  {France}, {Haswell}, {Ingrassia}, {Jenkins}, {Koskinen}, {Lanza}, {Redfield},
  {Youngblood}, \& {Pelzmann}}]{Fossati17}
{Fossati}, L., {Marcelja}, S.~E., {Staab}, D., {et~al.} 2017, ArXiv e-prints,
  arXiv:1702.02883

\bibitem[{{Fr{\"o}hlich} \& {Lean}(1998)}]{Frohlich98}
{Fr{\"o}hlich}, C., \& {Lean}, J. 1998, \grl, 25, 4377

\bibitem[{{Fr{\"o}hlich} \& {Lean}(2004)}]{Frohlich04}
---. 2004, \aapr, 12, 273

\bibitem[{{Gaia Collaboration} {et~al.}(2016){Gaia Collaboration}, {Prusti},
  {de Bruijne}, {Brown}, {Vallenari}, {Babusiaux}, {Bailer-Jones}, {Bastian},
  {Biermann}, {Evans}, \& et~al.}]{Prusti16}
{Gaia Collaboration}, {Prusti}, T., {de Bruijne}, J.~H.~J., {et~al.} 2016,
  \aap, 595, A1

\bibitem[{{Gaidos} {et~al.}(2016){Gaidos}, {Mann}, \& {Ansdell}}]{Gaidos16}
{Gaidos}, E., {Mann}, A.~W., \& {Ansdell}, M. 2016, \apj, 817, 50

\bibitem[{{Galilei} {et~al.}(1613){Galilei}, {Welser}, \& {de
  Filiis}}]{Galilei13}
{Galilei}, G., {Welser}, M., \& {de Filiis}, A. 1613, {Istoria E dimostrazioni
  intorno alle macchie solari E loro accidenti comprese in tre lettere scritte
  all'illvstrissimo signor Marco Velseri ...}

\bibitem[{{Garc{\'{\i}}a} {et~al.}(2010){Garc{\'{\i}}a}, {Mathur}, {Salabert},
  {Ballot}, {R{\'e}gulo}, {Metcalfe}, \& {Baglin}}]{Garcia10}
{Garc{\'{\i}}a}, R.~A., {Mathur}, S., {Salabert}, D., {et~al.} 2010, Science,
  329, 1032

\bibitem[{{Gilliland} {et~al.}(2010){Gilliland}, {Brown},
  {Christensen-Dalsgaard}, {Kjeldsen}, {Aerts}, {Appourchaux}, {Basu},
  {Bedding}, {Chaplin}, {Cunha}, {De Cat}, {De Ridder}, {Guzik}, {Handler},
  {Kawaler}, {Kiss}, {Kolenberg}, {Kurtz}, {Metcalfe}, {Monteiro}, {Szab{\'o}},
  {Arentoft}, {Balona}, {Debosscher}, {Elsworth}, {Quirion}, {Stello},
  {Su{\'a}rez}, {Borucki}, {Jenkins}, {Koch}, {Kondo}, {Latham}, {Rowe}, \&
  {Steffen}}]{Gilliland10}
{Gilliland}, R.~L., {Brown}, T.~M., {Christensen-Dalsgaard}, J., {et~al.} 2010,
  \pasp, 122, 131

\bibitem[{{Gilliland} {et~al.}(2011){Gilliland}, {Chaplin}, {Dunham},
  {Argabright}, {Borucki}, {Basri}, {Bryson}, {Buzasi}, {Caldwell}, {Elsworth},
  {Jenkins}, {Koch}, {Kolodziejczak}, {Miglio}, {van Cleve}, {Walkowicz}, \&
  {Welsh}}]{Gilliland11}
{Gilliland}, R.~L., {Chaplin}, W.~J., {Dunham}, E.~W., {et~al.} 2011, \apjs,
  197, 6

\bibitem[{{Gosling}(1993)}]{Gosling93}
{Gosling}, J.~T. 1993, \jgr, 98, 18937

\bibitem[{{Guo} {et~al.}(2017){Guo}, {Johnson}, {Mann}, {Kraus}, {Curtis}, \&
  {Latham}}]{Guo17}
{Guo}, X., {Johnson}, J.~A., {Mann}, A.~W., {et~al.} 2017, \apj, 838, 25

\bibitem[{{Haas} {et~al.}(2010){Haas}, {Batalha}, {Bryson}, {Caldwell},
  {Dotson}, {Hall}, {Jenkins}, {Klaus}, {Koch}, {Kolodziejczak}, {Middour},
  {Smith}, {Sobeck}, {Stober}, {Thompson}, \& {Van Cleve}}]{Haas10}
{Haas}, M.~R., {Batalha}, N.~M., {Bryson}, S.~T., {et~al.} 2010, \apjl, 713,
  L115

\bibitem[{{Hall} {et~al.}(2009){Hall}, {Henry}, {Lockwood}, {Skiff}, \&
  {Saar}}]{Hall09}
{Hall}, J.~C., {Henry}, G.~W., {Lockwood}, G.~W., {Skiff}, B.~A., \& {Saar},
  S.~H. 2009, \aj, 138, 312

\bibitem[{{Hathaway}(2015)}]{Hathaway15}
{Hathaway}, D.~H. 2015, Living Reviews in Solar Physics, 12, 4

\bibitem[{{Hawley} {et~al.}(2014){Hawley}, {Davenport}, {Kowalski},
  {Wisniewski}, {Hebb}, {Deitrick}, \& {Hilton}}]{Hawley14}
{Hawley}, S.~L., {Davenport}, J.~R.~A., {Kowalski}, A.~F., {et~al.} 2014, \apj,
  797, 121

\bibitem[{{Huber} {et~al.}(2013){Huber}, {Carter}, {Barbieri}, {Miglio},
  {Deck}, {Fabrycky}, {Montet}, {Buchhave}, {Chaplin}, {Hekker},
  {Montalb{\'a}n}, {Sanchis-Ojeda}, {Basu}, {Bedding}, {Campante},
  {Christensen-Dalsgaard}, {Elsworth}, {Stello}, {Arentoft}, {Ford},
  {Gilliland}, {Handberg}, {Howard}, {Isaacson}, {Johnson}, {Karoff},
  {Kawaler}, {Kjeldsen}, {Latham}, {Lund}, {Lundkvist}, {Marcy}, {Metcalfe},
  {Silva Aguirre}, \& {Winn}}]{Huber13}
{Huber}, D., {Carter}, J.~A., {Barbieri}, M., {et~al.} 2013, Science, 342, 331

\bibitem[{{Huber} {et~al.}(2014){Huber}, {Silva Aguirre}, {Matthews},
  {Pinsonneault}, {Gaidos}, {Garc{\'{\i}}a}, {Hekker}, {Mathur}, {Mosser},
  {Torres}, {Bastien}, {Basu}, {Bedding}, {Chaplin}, {Demory}, {Fleming},
  {Guo}, {Mann}, {Rowe}, {Serenelli}, {Smith}, \& {Stello}}]{Huber14}
{Huber}, D., {Silva Aguirre}, V., {Matthews}, J.~M., {et~al.} 2014, \apjs, 211,
  2

\bibitem[{Hunter {et~al.}(2007)}]{matplotlib}
Hunter, J.~D., {et~al.} 2007, Computing in science and engineering, 9, 90

\bibitem[{{Ibano{\v g}lu} {et~al.}(2001){Ibano{\v g}lu}, {Evren}, {Ta{\v s}},
  {Devlen}, \& {{\c C}ak{\"i}rl{\"i}}}]{Ibanoglu01}
{Ibano{\v g}lu}, C., {Evren}, S., {Ta{\v s}}, G., {Devlen}, A., \& {{\c
  C}ak{\"i}rl{\"i}}, {\"O}. 2001, \aap, 371, 626

\bibitem[{{Jenkins} {et~al.}(2010){Jenkins}, {Caldwell}, {Chandrasekaran},
  {Twicken}, {Bryson}, {Quintana}, {Clarke}, {Li}, {Allen}, {Tenenbaum}, {Wu},
  {Klaus}, {Middour}, {Cote}, {McCauliff}, {Girouard}, {Gunter}, {Wohler},
  {Sommers}, {Hall}, {Uddin}, {Wu}, {Bhavsar}, {Van Cleve}, {Pletcher},
  {Dotson}, {Haas}, {Gilliland}, {Koch}, \& {Borucki}}]{Jenkins10}
{Jenkins}, J.~M., {Caldwell}, D.~A., {Chandrasekaran}, H., {et~al.} 2010,
  \apjl, 713, L87

\bibitem[{{Jenkins} {et~al.}(2015){Jenkins}, {Twicken}, {Batalha}, {Caldwell},
  {Cochran}, {Endl}, {Latham}, {Esquerdo}, {Seader}, {Bieryla}, {Petigura},
  {Ciardi}, {Marcy}, {Isaacson}, {Huber}, {Rowe}, {Torres}, {Bryson},
  {Buchhave}, {Ramirez}, {Wolfgang}, {Li}, {Campbell}, {Tenenbaum},
  {Sanderfer}, {Henze}, {Catanzarite}, {Gilliland}, \& {Borucki}}]{Jenkins15}
{Jenkins}, J.~M., {Twicken}, J.~D., {Batalha}, N.~M., {et~al.} 2015, \aj, 150,
  56

\bibitem[{{Jordi} {et~al.}(2010){Jordi}, {Gebran}, {Carrasco}, {de Bruijne},
  {Voss}, {Fabricius}, {Knude}, {Vallenari}, {Kohley}, \& {Mora}}]{Jordi10}
{Jordi}, C., {Gebran}, M., {Carrasco}, J.~M., {et~al.} 2010, \aap, 523, A48

\bibitem[{Kingma \& Ba(2014)}]{Kingma14}
Kingma, D.~P., \& Ba, J. 2014, in Proceedings of the 3rd International
  Conference on Learning Representations (ICLR)

\bibitem[{{Lehmann} {et~al.}(2012){Lehmann}, {Zechmeister}, {Dreizler},
  {Schuh}, \& {Kanzler}}]{Lehmann12}
{Lehmann}, H., {Zechmeister}, M., {Dreizler}, S., {Schuh}, S., \& {Kanzler}, R.
  2012, \aap, 541, A105

\bibitem[{{Libralato} {et~al.}(2016){Libralato}, {Bedin}, {Nardiello}, \&
  {Piotto}}]{Libralato16}
{Libralato}, M., {Bedin}, L.~R., {Nardiello}, D., \& {Piotto}, G. 2016, \mnras,
  456, 1137

\bibitem[{{Lockwood} {et~al.}(1997){Lockwood}, {Skiff}, \&
  {Radick}}]{Lockwood97}
{Lockwood}, G.~W., {Skiff}, B.~A., \& {Radick}, R.~R. 1997, \apj, 485, 789

\bibitem[{{Luger} {et~al.}(2016){Luger}, {Agol}, {Kruse}, {Barnes}, {Becker},
  {Foreman-Mackey}, \& {Deming}}]{Luger16}
{Luger}, R., {Agol}, E., {Kruse}, E., {et~al.} 2016, \aj, 152, 100

\bibitem[{{Mamajek} \& {Hillenbrand}(2008)}]{Mamajek08}
{Mamajek}, E.~E., \& {Hillenbrand}, L.~A. 2008, \apj, 687, 1264

\bibitem[{{Marsh} {et~al.}(2017){Marsh}, {Prince}, {Mahabal}, {Bellm}, {Drake},
  \& {Djorgovski}}]{Marsh17}
{Marsh}, F.~M., {Prince}, T.~A., {Mahabal}, A.~A., {et~al.} 2017, \mnras, 465,
  4678

\bibitem[{{Mathur} {et~al.}(2014){Mathur}, {Garc{\'{\i}}a}, {Ballot},
  {Ceillier}, {Salabert}, {Metcalfe}, {R{\'e}gulo}, {Jim{\'e}nez}, \&
  {Bloemen}}]{Mathur14}
{Mathur}, S., {Garc{\'{\i}}a}, R.~A., {Ballot}, J., {et~al.} 2014, \aap, 562,
  A124

\bibitem[{{McQuillan} {et~al.}(2014){McQuillan}, {Mazeh}, \&
  {Aigrain}}]{McQuillan14}
{McQuillan}, A., {Mazeh}, T., \& {Aigrain}, S. 2014, \apjs, 211, 24

\bibitem[{{Metcalfe} {et~al.}(2010){Metcalfe}, {Basu}, {Henry}, {Soderblom},
  {Judge}, {Kn{\"o}lker}, {Mathur}, \& {Rempel}}]{Metcalfe10}
{Metcalfe}, T.~S., {Basu}, S., {Henry}, T.~J., {et~al.} 2010, \apjl, 723, L213

\bibitem[{{Metcalfe} {et~al.}(2016){Metcalfe}, {Egeland}, \& {van
  Saders}}]{Metcalfe16}
{Metcalfe}, T.~S., {Egeland}, R., \& {van Saders}, J. 2016, \apjl, 826, L2

\bibitem[{{Metcalfe} {et~al.}(2013){Metcalfe}, {Buccino}, {Brown}, {Mathur},
  {Soderblom}, {Henry}, {Mauas}, {Petrucci}, {Hall}, \& {Basu}}]{Metcalfe13}
{Metcalfe}, T.~S., {Buccino}, A.~P., {Brown}, B.~P., {et~al.} 2013, \apjl, 763,
  L26

\bibitem[{{Meyer} \& {Blades}(1996)}]{Meyer96}
{Meyer}, D.~M., \& {Blades}, J.~C. 1996, \apjl, 464, L179

\bibitem[{{Molnar} {et~al.}(2017){Molnar}, {Van Noord}, {Kinemuchi},
  {Smolinski}, {Alexander}, {Cook}, {Jang}, {Kobulnicky}, {Spedden}, \&
  {Steenwyk}}]{Molnar17}
{Molnar}, L.~A., {Van Noord}, D.~M., {Kinemuchi}, K., {et~al.} 2017, \apj, 840,
  1

\bibitem[{{Montet} \& {Simon}(2016)}]{Montet16b}
{Montet}, B.~T., \& {Simon}, J.~D. 2016, \apjl, 830, L39

\bibitem[{{Mullally} {et~al.}(2016){Mullally}, {Barclay}, \&
  {Barentsen}}]{Mullally16}
{Mullally}, F., {Barclay}, T., \& {Barentsen}, G. 2016, {K2fov: Field of view
  software for NASA's K2 mission}, Astrophysics Source Code Library, , ,
  ascl:1601.009

\bibitem[{{Nielsen} {et~al.}(2013){Nielsen}, {Gizon}, {Schunker}, \&
  {Karoff}}]{Nielsen13}
{Nielsen}, M.~B., {Gizon}, L., {Schunker}, H., \& {Karoff}, C. 2013, \aap, 557,
  L10

\bibitem[{{Noyes} {et~al.}(1984){Noyes}, {Hartmann}, {Baliunas}, {Duncan}, \&
  {Vaughan}}]{Noyes84}
{Noyes}, R.~W., {Hartmann}, L.~W., {Baliunas}, S.~L., {Duncan}, D.~K., \&
  {Vaughan}, A.~H. 1984, \apj, 279, 763

\bibitem[{{Ol{\'a}h} {et~al.}(2016){Ol{\'a}h}, {K{\H o}v{\'a}ri}, {Petrovay},
  {Soon}, {Baliunas}, {Koll{\'a}th}, \& {Vida}}]{Olah16}
{Ol{\'a}h}, K., {K{\H o}v{\'a}ri}, Z., {Petrovay}, K., {et~al.} 2016, \aap,
  590, A133

\bibitem[{{Olmedo} {et~al.}(2015){Olmedo}, {Lloyd}, {Mamajek}, {Ch{\'a}vez},
  {Bertone}, {Martin}, \& {Neill}}]{Olmedo15}
{Olmedo}, M., {Lloyd}, J., {Mamajek}, E.~E., {et~al.} 2015, \apj, 813, 100

\bibitem[{{Padmanabhan} {et~al.}(2008){Padmanabhan}, {Schlegel}, {Finkbeiner},
  {Barentine}, {Blanton}, {Brewington}, {Gunn}, {Harvanek}, {Hogg},
  {Ivezi{\'c}}, {Johnston}, {Kent}, {Kleinman}, {Knapp}, {Krzesinski}, {Long},
  {Neilsen}, {Nitta}, {Loomis}, {Lupton}, {Roweis}, {Snedden}, {Strauss}, \&
  {Tucker}}]{Padmanabhan08}
{Padmanabhan}, N., {Schlegel}, D.~J., {Finkbeiner}, D.~P., {et~al.} 2008, \apj,
  674, 1217

\bibitem[{{Pinsonneault} {et~al.}(2012){Pinsonneault}, {An},
  {Molenda-{\.Z}akowicz}, {Chaplin}, {Metcalfe}, \& {Bruntt}}]{Pinsonneault12}
{Pinsonneault}, M.~H., {An}, D., {Molenda-{\.Z}akowicz}, J., {et~al.} 2012,
  \apjs, 199, 30

\bibitem[{{Quintana} {et~al.}(2010){Quintana}, {Jenkins}, {Clarke},
  {Chandrasekaran}, {Twicken}, {McCauliff}, {Cote}, {Klaus}, {Allen},
  {Caldwell}, \& {Bryson}}]{Quintana10}
{Quintana}, E.~V., {Jenkins}, J.~M., {Clarke}, B.~D., {et~al.} 2010, in
  \procspie, Vol. 7740, Software and Cyberinfrastructure for Astronomy, 77401X

\bibitem[{{Radick} {et~al.}(1998){Radick}, {Lockwood}, {Skiff}, \&
  {Baliunas}}]{Radick98}
{Radick}, R.~R., {Lockwood}, G.~W., {Skiff}, B.~A., \& {Baliunas}, S.~L. 1998,
  \apjs, 118, 239

\bibitem[{{Ram{\'{\i}}rez} {et~al.}(2012){Ram{\'{\i}}rez}, {Michel}, {Sefako},
  {Tucci Maia}, {Schuster}, {van Wyk}, {Mel{\'e}ndez}, {Casagrande}, \&
  {Castilho}}]{Ramirez12}
{Ram{\'{\i}}rez}, I., {Michel}, R., {Sefako}, R., {et~al.} 2012, \apj, 752, 5

\bibitem[{{Rauer} {et~al.}(2014){Rauer}, {Catala}, {Aerts}, {Appourchaux},
  {Benz}, {Brandeker}, {Christensen-Dalsgaard}, {Deleuil}, {Gizon}, {Goupil},
  {G{\"u}del}, {Janot-Pacheco}, {Mas-Hesse}, {Pagano}, {Piotto}, {Pollacco},
  {Santos}, {Smith}, {Su{\'a}rez}, {Szab{\'o}}, {Udry}, {Adibekyan}, {Alibert},
  {Almenara}, {Amaro-Seoane}, {Eiff}, {Asplund}, {Antonello}, {Barnes},
  {Baudin}, {Belkacem}, {Bergemann}, {Bihain}, {Birch}, {Bonfils}, {Boisse},
  {Bonomo}, {Borsa}, {Brand{\~a}o}, {Brocato}, {Brun}, {Burleigh}, {Burston},
  {Cabrera}, {Cassisi}, {Chaplin}, {Charpinet}, {Chiappini}, {Church},
  {Csizmadia}, {Cunha}, {Damasso}, {Davies}, {Deeg}, {D{\'{\i}}az}, {Dreizler},
  {Dreyer}, {Eggenberger}, {Ehrenreich}, {Eigm{\"u}ller}, {Erikson}, {Farmer},
  {Feltzing}, {de Oliveira Fialho}, {Figueira}, {Forveille}, {Fridlund},
  {Garc{\'{\i}}a}, {Giommi}, {Giuffrida}, {Godolt}, {Gomes da Silva},
  {Granzer}, {Grenfell}, {Grotsch-Noels}, {G{\"u}nther}, {Haswell}, {Hatzes},
  {H{\'e}brard}, {Hekker}, {Helled}, {Heng}, {Jenkins}, {Johansen},
  {Khodachenko}, {Kislyakova}, {Kley}, {Kolb}, {Krivova}, {Kupka}, {Lammer},
  {Lanza}, {Lebreton}, {Magrin}, {Marcos-Arenal}, {Marrese}, {Marques},
  {Martins}, {Mathis}, {Mathur}, {Messina}, {Miglio}, {Montalban}, {Montalto},
  {Monteiro}, {Moradi}, {Moravveji}, {Mordasini}, {Morel}, {Mortier},
  {Nascimbeni}, {Nelson}, {Nielsen}, {Noack}, {Norton}, {Ofir}, {Oshagh},
  {Ouazzani}, {P{\'a}pics}, {Parro}, {Petit}, {Plez}, {Poretti}, {Quirrenbach},
  {Ragazzoni}, {Raimondo}, {Rainer}, {Reese}, {Redmer}, {Reffert},
  {Rojas-Ayala}, {Roxburgh}, {Salmon}, {Santerne}, {Schneider}, {Schou},
  {Schuh}, {Schunker}, {Silva-Valio}, {Silvotti}, {Skillen}, {Snellen}, {Sohl},
  {Sousa}, {Sozzetti}, {Stello}, {Strassmeier}, {{\v S}vanda}, {Szab{\'o}},
  {Tkachenko}, {Valencia}, {Van Grootel}, {Vauclair}, {Ventura}, {Wagner},
  {Walton}, {Weingrill}, {Werner}, {Wheatley}, \& {Zwintz}}]{Rauer14}
{Rauer}, H., {Catala}, C., {Aerts}, C., {et~al.} 2014, Experimental Astronomy,
  38, 249

\bibitem[{{Reinhold} {et~al.}(2017){Reinhold}, {Cameron}, \&
  {Gizon}}]{Reinhold17}
{Reinhold}, T., {Cameron}, R.~H., \& {Gizon}, L. 2017, ArXiv e-prints,
  arXiv:1705.03312

\bibitem[{{Ricker} {et~al.}(2014){Ricker}, {Winn}, {Vanderspek}, {Latham},
  {Bakos}, {Bean}, {Berta-Thompson}, {Brown}, {Buchhave}, {Butler}, {Butler},
  {Chaplin}, {Charbonneau}, {Christensen-Dalsgaard}, {Clampin}, {Deming},
  {Doty}, {De Lee}, {Dressing}, {Dunham}, {Endl}, {Fressin}, {Ge}, {Henning},
  {Holman}, {Howard}, {Ida}, {Jenkins}, {Jernigan}, {Johnson}, {Kaltenegger},
  {Kawai}, {Kjeldsen}, {Laughlin}, {Levine}, {Lin}, {Lissauer}, {MacQueen},
  {Marcy}, {McCullough}, {Morton}, {Narita}, {Paegert}, {Palle}, {Pepe},
  {Pepper}, {Quirrenbach}, {Rinehart}, {Sasselov}, {Sato}, {Seager},
  {Sozzetti}, {Stassun}, {Sullivan}, {Szentgyorgyi}, {Torres}, {Udry}, \&
  {Villasenor}}]{Ricker14}
{Ricker}, G.~R., {Winn}, J.~N., {Vanderspek}, R., {et~al.} 2014, in \procspie,
  Vol. 9143, Space Telescopes and Instrumentation 2014: Optical, Infrared, and
  Millimeter Wave, 914320

\bibitem[{{Saar} \& {Brandenburg}(1999)}]{Saar99}
{Saar}, S.~H., \& {Brandenburg}, A. 1999, \apj, 524, 295

\bibitem[{{Salabert} {et~al.}(2016{\natexlab{a}}){Salabert}, {R{\'e}gulo},
  {Garc{\'{\i}}a}, {Beck}, {Ballot}, {Creevey}, {P{\'e}rez Hern{\'a}ndez}, {do
  Nascimento}, {Corsaro}, {Egeland}, {Mathur}, {Metcalfe}, {Bigot}, {Ceillier},
  \& {Pall{\'e}}}]{Salabert16a}
{Salabert}, D., {R{\'e}gulo}, C., {Garc{\'{\i}}a}, R.~A., {et~al.}
  2016{\natexlab{a}}, \aap, 589, A118

\bibitem[{{Salabert} {et~al.}(2016{\natexlab{b}}){Salabert}, {Garc{\'{\i}}a},
  {Beck}, {Egeland}, {Pall{\'e}}, {Mathur}, {Metcalfe}, {do Nascimento},
  {Ceillier}, {Andersen}, \& {Trivi{\~n}o Hage}}]{Salabert16b}
{Salabert}, D., {Garc{\'{\i}}a}, R.~A., {Beck}, P.~G., {et~al.}
  2016{\natexlab{b}}, \aap, 596, A31

\bibitem[{{Schwabe}(1844)}]{Schwabe44}
{Schwabe}, M. 1844, Astronomische Nachrichten, 21, 233

\bibitem[{{Shapiro} {et~al.}(2014){Shapiro}, {Solanki}, {Krivova}, {Schmutz},
  {Ball}, {Knaack}, {Rozanov}, \& {Unruh}}]{Shapiro14}
{Shapiro}, A.~I., {Solanki}, S.~K., {Krivova}, N.~A., {et~al.} 2014, \aap, 569,
  A38

\bibitem[{{Silva Aguirre} {et~al.}(2015){Silva Aguirre}, {Davies}, {Basu},
  {Christensen-Dalsgaard}, {Creevey}, {Metcalfe}, {Bedding}, {Casagrande},
  {Handberg}, {Lund}, {Nissen}, {Chaplin}, {Huber}, {Serenelli}, {Stello}, {Van
  Eylen}, {Campante}, {Elsworth}, {Gilliland}, {Hekker}, {Karoff}, {Kawaler},
  {Kjeldsen}, \& {Lundkvist}}]{Aguirre15}
{Silva Aguirre}, V., {Davies}, G.~R., {Basu}, S., {et~al.} 2015, \mnras, 452,
  2127

\bibitem[{{Smith} {et~al.}(2012){Smith}, {Stumpe}, {Van Cleve}, {Jenkins},
  {Barclay}, {Fanelli}, {Girouard}, {Kolodziejczak}, {McCauliff}, {Morris}, \&
  {Twicken}}]{Smith12}
{Smith}, J.~C., {Stumpe}, M.~C., {Van Cleve}, J.~E., {et~al.} 2012, \pasp, 124,
  1000

\bibitem[{{Soderblom}(2010)}]{Soderblom10}
{Soderblom}, D.~R. 2010, \araa, 48, 581

\bibitem[{{Stumpe} {et~al.}(2012){Stumpe}, {Smith}, {Van Cleve}, {Twicken},
  {Barclay}, {Fanelli}, {Girouard}, {Jenkins}, {Kolodziejczak}, {McCauliff}, \&
  {Morris}}]{Stumpe12}
{Stumpe}, M.~C., {Smith}, J.~C., {Van Cleve}, J.~E., {et~al.} 2012, \pasp, 124,
  985

\bibitem[{{Theano Development Team}(2016)}]{Theano}
{Theano Development Team}. 2016, arXiv e-prints, abs/1605.02688.
\newblock \url{http://arxiv.org/abs/1605.02688}

\bibitem[{{Tregloan-Reed} \& {Southworth}(2013)}]{TregloanReed13}
{Tregloan-Reed}, J., \& {Southworth}, J. 2013, \mnras, 431, 966

\bibitem[{{Tylenda} {et~al.}(2011){Tylenda}, {Hajduk}, {Kami{\'n}ski},
  {Udalski}, {Soszy{\'n}ski}, {Szyma{\'n}ski}, {Kubiak}, {Pietrzy{\'n}ski},
  {Poleski}, {Wyrzykowski}, \& {Ulaczyk}}]{Tylenda11}
{Tylenda}, R., {Hajduk}, M., {Kami{\'n}ski}, T., {et~al.} 2011, \aap, 528, A114

\bibitem[{{Van Cleve} {et~al.}(2016){Van Cleve}, {Howell}, {Smith}, {Clarke},
  {Thompson}, {Bryson}, {Lund}, {Handberg}, \& {Chaplin}}]{VanCleve16}
{Van Cleve}, J.~E., {Howell}, S.~B., {Smith}, J.~C., {et~al.} 2016, \pasp, 128,
  075002

\bibitem[{Van Der~Walt {et~al.}(2011)Van Der~Walt, Colbert, \&
  Varoquaux}]{numpy}
Van Der~Walt, S., Colbert, S.~C., \& Varoquaux, G. 2011, Computing in Science
  \& Engineering, 13, 22

\bibitem[{{van Saders} {et~al.}(2016){van Saders}, {Ceillier}, {Metcalfe},
  {Silva Aguirre}, {Pinsonneault}, {Garc{\'{\i}}a}, {Mathur}, \&
  {Davies}}]{VanSaders16}
{van Saders}, J.~L., {Ceillier}, T., {Metcalfe}, T.~S., {et~al.} 2016, \nat,
  529, 181

\bibitem[{{Wilson}(1978)}]{Wilson78}
{Wilson}, O.~C. 1978, \apj, 226, 379

\end{thebibliography}
\end{document}